\documentclass[aps,prb,twocolumn,superscriptaddress,showpacs,preprintnumbers]{revtex4-2}
\usepackage[T1]{fontenc}
\usepackage[utf8x]{inputenc}
\usepackage{gensymb}
\usepackage{times}
\usepackage{color}
\usepackage[colorlinks,bookmarks=false,citecolor=blue,linkcolor=red,urlcolor=blue]{hyperref}
\usepackage{colortbl,amsthm,amsmath,amssymb,txfonts}
\usepackage{graphicx}
\usepackage{bm}
\usepackage{ragged2e}

\usepackage{appendix}

\begin{document}

\title{Critical Behavior and Collective Modes at the Superfluid Transition in Amorphous Systems }

\author{Pulloor Kuttanikkad Vishnu}
 \affiliation{Department of Physics, Indian Institute of Technology Madras, Chennai 600036, India.}
 \author{Martin Puschmann}
 \affiliation{Department of Physics, Missouri University of Science and Technology, Rolla, Missouri 65409, USA.}
\author{Rajesh Narayanan}
\affiliation{Department of Physics, Indian Institute of Technology Madras, Chennai 600036, India.}
 \email{rnarayanan@iitm.ac.in }
 \author{Thomas Vojta}
\affiliation{Department of Physics, Missouri University of Science and Technology, Rolla, Missouri 65409, USA.}

\begin{abstract}

We investigate the critical behavior and the dynamics of the amplitude (Higgs) mode close to the superfluid-insulator quantum phase transition in an amorphous system (i.e., a system subject to topological randomness). In particular, we
map the two-dimensional Bose-Hubbard Hamiltonian defined on a random Voronoi-Delaunay lattice onto a (2+1)-dimensional layered classical XY model with correlated topological disorder. We study the resulting model by laying recourse to classical Monte Carlo simulations. We specifically focus on the scalar susceptibility of the order parameter to study the dynamics of the amplitude mode. To do so, we harness the maximum entropy method to perform the analytic continuation of the scalar susceptibility to real frequencies. Our analysis shows that the amplitude mode remains delocalized in the presence of such topological disorder, quite at odds with its behavior in generic disordered systems, where the randomness localizes the Higgs mode. Furthermore, we show that the critical behavior of the topologically disordered system is identical to that of its translationally invariant counterpart, consistent with a modified Harris criterion. This suggests that the localization of the collective excitations in the presence of disorder is tied to the critical behavior of the quantum phase transition rather than a simple Anderson-localization-type interference mechanism.
\end{abstract}

\date{\today}
\maketitle

\section{Introduction}
\label{sec1}

It is well established that quantum phase transitions in many-body systems are deeply impacted by quenched disorder. Research over the past few decades has shown that random disorder typified by vacancies, defects, and other types of impurities can trigger a plethora of exciting novel phenomena such as quantum Griffiths phases \cite{GuoBhattHuse96,RiegerYoung96,VojtaSchmalian05}, infinite-randomness criticality \cite{Fisher92,*Fisher95,hoyos_kotabage_prl_07,*vojta_kotabage_prb_09}, other unconventional critical points \cite{AKPR04,*AKPR08,vojta_hoyos_jpcm_11}, as well as smeared transitions \cite{Vojta03a,*HoyosVojta08}. Classification schemes \cite{MMHF00,VojtaHoyos14} for these phenomena have been put forward according to the behavior of the disorder
strength under coarse graining and on the importance of rare disorder fluctuations (for reviews see, e.g., Refs.\ \cite{vojta_jpamg_06,*Vojta13,*Vojta19}).
Many of these exotic phenomena have now been realized experimentally \cite{ubaid_vojta_prl_10,shi_lin_natphy_14,xing_zhang_sci_15,shen_xing_prb_16,xing_zhao_nanlet_17,wang_gebretsadik_prl_17,saito_nojima_natcom_18,zhang_fan_npgam_19,lewellyn_percher_prb_19,reiss_graf_prl_21,kaur_kundu_arxiv_22}.

The majority of the investigations in the literature on disorder effects at quantum phase transitions have been restricted to the thermodynamic behavior. Only recently, attention has focussed on the effects of impurities on the dynamics of excitations hosted by many-body systems near quantum criticality. Of particular interest is the role that impurities play for the properties of the amplitude (Higgs) modes and the Goldstone modes that emerge when a continuous symmetry is spontaneously broken.

Indeed, the question of disorder effects on the amplitude modes has gathered a renewed impetus, as these modes can now be experimentally realized in various magnetic, superconducting, cold atom, and charge-density-wave systems \cite{shimano_tsuji_arcmp_20, pekker_varma_arcmp_15}. In the case of clean systems, amplitude modes can be identified as sharp excitations in the symmetry-broken phase, close to the critical point \cite{podolsky_auerbach_prb_11,podolsky_sachdev_prb_12,gazit_podolsky_prl_13}. More recently, a number of studies investigating the effects of disorder on collective excitation have come to the fore. A series of papers \cite{puschmann_crewse_prl_20, crewse_vojta_prb_21,puschmann_getelina_annals_21} has revealed that the amplitude mode obtained at a superfluid-Mott glass transition is localized by the quenched impurities. Furthermore, it has been shown that, in accordance with the Goldstone theorem, only the lowest-lying Goldstone mode remains delocalized.

Even though there is ample evidence suggesting the localization of collective modes, the mechanism underlying the localization is not fully resolved. Does it hinge on a simple Anderson-localization type interference mechanism \cite{anderson_pr_58} or is it tied to the critical behavior of the underlying quantum phase transition via mode coupling effects?

To shed light on this question, we study the collective excitations, particularly the amplitude mode near the superfluid-insulator transition on a random Voronoi-Delaunay (VD) lattice \cite{okabe_boots_book_00}, where the disorder is exemplified by the random connectivity of the lattice sites. (This type of disorder is often called topological disorder.) Random VD lattices are commonly found in nature in the form of amorphous materials, foams, animal skins, muscle fibers, and plant tissues and are used in different applications ranging from material science to astronomy \cite{weaire_rivier_cp_84, franz_acm_91, okabe_boots_book_00}. It is now known that the Harris criterion \cite{Harris74} gets modified on such topological lattices because the disorder displays strong anti-correlations. It thus decays qualitatively faster with increasing length scale than generic quenched disorder \cite{barghathi_vojta_prl_14}. The relevancy of the disorder operator is now set by a modified criterion which states that in the case of topological disorder, the clean critical behavior is destabilized by impurity fluctuations if the inequality  $(d+1)\nu>2$ is violated. Here $\nu$ is the correlation length critical exponent and $d$ is the space dimensionality of the system. It is well known, for instance, that the random connectivity disorder is irrelevant at the critical point of the 3D Ising model \cite{janke_villanova_prb_02, lima_costa_physica_08}, in accordance with this modified Harris criterion. In contrast, it was shown in Ref. \cite{puschmann_cain_epjb_15} that Anderson localization on a two-dimensional random VD lattice is not affected qualitatively by the disorder anticorrelations and thus behaves analogously to the case of generic randomness, i.e., all states are localized. This implies that analyzing the amplitude mode on a random VD lattice offers a route to disentangle the possible localization mechanisms.

In this paper, we therefore study a system of interacting bosons on a random VD lattice across the superfluid-insulator quantum phase transition. The purpose is twofold. First, we want to determine the quantum critical behavior to verify that this system is indeed governed by the modified Harris criterion \cite{barghathi_vojta_prl_14} rather than the regular one. Second, we want to analyze the collective excitations in the superfluid phase and their localization properties. To do so, we employ a quantum-to-classical mapping to rewrite the Bose-Hubbard model (in the particle-hole symmetric limit) onto a classical (2+1) dimensional XY model \cite{wallin_sorensen_prb_94}. The mapped classical model is studied using large-scale Monte Carlo simulations. The scalar susceptibility of the order parameter is obtained in terms of imaginary frequency and then converted to real frequency using the maximum entropy method (MaxEnt) \cite{jarrell_guernatis_pr_96}. In accordance with the modified Harris criterion, we find that the critical behavior of the topologically disordered system is identical to that of its clean, translationally invariant counterpart. In contrast, the superfluid-Mott glass quantum phase transition in a generically disordered system was shown to be in a novel universality class \cite{vojta_crewse_prb_16}.
We further show that the amplitude mode remains delocalized across the transition of the topologically disordered system, while it was shown to be spatially localized for generic disorder \cite{puschmann_crewse_prl_20,crewse_vojta_prb_21,puschmann_getelina_annals_21}. This suggests that the localization of the amplitude mode is caused by mode-coupling and renormalization effects and tied to the critical behavior of the transition rather than a simple Anderson-localization mechanism (because, in the latter case, one would expect localization on a random VD lattice).

The rest of this paper is organized as follows. In Section \ref{sec2}, we discuss the model Hamiltonian and its mapping to the corresponding classical system. In Section \ref{sec3}, we discuss the methods used in our study, including the construction of the random VD lattice, the classical Monte Carlo simulations, and the maximum entropy method. The critical behavior of the system is described in detail in Section \ref{sec4}. The same section also addresses the localization properties of the amplitude modes. A semi-analytical inhomogeneous mean-field analysis in support of our study is described in Section \ref{sec5}. Section \ref{sec6} is devoted to the effects of generic randomness on top of the topological disorder of the random VD lattice. We conclude our findings in Section \ref{sec7}.

\section{Model}
\label{sec2}

Our starting point is the Bose-Hubbard model on a two-dimensional random VD lattice. This lattice consists of a set of lattice sites at independent random positions, connected by bonds that are defined by means of the VD construction \cite{okabe_boots_book_00}. Details of the lattice construction will be given in Sec.\ \ref{sec3-Voronoi}. The Bose-Hubbard Hamiltonian is given by:
\begin{equation}
    H=\frac{1}{2}\sum_{i}U_{i}(\hat{n}_{i}-\tilde{n})^{2}-\sum_{<ij>}J_{ij}(a_{i}^{\dagger} a_{j} + \textrm{h.c.})~.
    \label{eqn.1}
\end{equation}
Here $a_{i}$, $a_{i}^{\dagger}$, $\hat{n}_{i}$ are boson annihilation, creation, and number operators, respectively. Further, $U_{i}$ is the Hubbard interaction at site $i$, and $\tilde{n}$ is the average filling which we fix at a large integer value to ensure particle-hole symmetry. Finally, $J_{ij}$ denotes the hopping amplitude between the nearest neighbor sites of the random VD lattice. For pure connectivity disorder, we set
$J_{ij}\equiv J$ for all nearest neighbors. Additional generic disorder can be introduced by site or bond dilutions or by drawing $J_{ij}$ from a random distribution.
The translationally invariant (clean) version of the model (\ref{eqn.1}), defined, e.g., on a square lattice, is known to host two phases, namely the Mott insulator and superfluid phases, separated by a quantum phase transition in the 3D XY universality class \cite{WeichmanMukhopadhyay08}.

\begin{figure}
    \centering
    \includegraphics[width=\columnwidth]{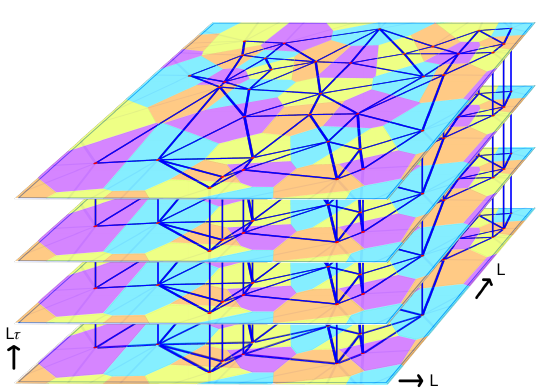}
    \caption{\justifying Layered (2+1)-dimensional random VD lattice. The same Voronoi layer is repeated in the vertical (imaginary-time) direction.}
    \label{fig:2+1_voronoi_lattice}
\end{figure}

In preparation for the Monte Carlo simulations, we rewrite the model (\ref{eqn.1}) by means of a quantum-to-classical mapping \cite{wallin_sorensen_prb_94, crewse_vojta_prb_21,puschmann_crewse_prl_20}
onto a layered (2+1)-dimensional classical XY model, see Fig.\ \ref{fig:2+1_voronoi_lattice}.
Here, the third dimension represents imaginary time. The classical Hamiltonian takes the form
\begin{equation}
    H_{\rm cl}=-\sum_{\left<i,j\right>,\tau}J^{s}_{ij}\mathbf{S}_{i,\tau}\cdot\mathbf{S}_{j,\tau}-J^{t}\sum_{i,\tau}\mathbf{S}_{i,\tau}\cdot\mathbf{S}_{i,\tau+1}~,
    \label{eqn.2}
\end{equation}
where $\mathbf{S}_{j,\tau}$ is a classical $O(2)$ rotor, i.e., a two-component unit vector, situated at spatial coordinate $i$ and imaginary time $\tau$. All layers consist of the same two-dimensional VD tessellation. The spatial interactions $J^{s}_{ij}$ couple Voronoi neighbors in the same layer (labeled by the layer or imaginary time index $\tau$). $J^{t}$ elucidates the coupling between equivalent sites in neighboring VD layers, i.e., in the imaginary time direction.
Due to the quantum-to-classical mapping, the disorder in the classical Hamiltonian is perfectly correlated in the imaginary-time dimension. This not only means that each layer contains the same VD tessellation, it also implies that the interactions  $J^{s}$ and  $J^{t}$ do not depend on the imaginary-time variable $\tau$.

The interaction constants appearing in the quantum and classical Hamiltonians are related via $\beta_{c}J^{s}\sim J$ and $\beta_{c}J^{t}\sim 1/U$, where $\beta_{c} = 1/T$ is the inverse temperature of the classical model (with the Boltzmann constant set to unity). This classical temperature does not correspond to the actual temperature of the quantum system, which remains at absolute zero. Thus, we can investigate the critical behavior of the superfluid-insulator quantum phase transition occurring at zero temperature as a function of the ratio $U/J$ by tuning the corresponding classical temperature $T$ in the classical Hamiltonian (\ref{eqn.2}). As we are interested in universal properties, we set $J^{t}$ as well as all $J^{s}_{ij}$ between Voronoi-Delaunay nearest neighbors to unity in the Monte Carlo simulations.

\section{Methods}
\label{sec3}
\subsection{Construction of the random Voronoi-Delaunay lattice}
\label{sec3-Voronoi}

The random VD lattice consists of a set of lattice sites at random positions uniformly distributed in a square box of linear size $L$ with periodic boundary conditions. The point density is fixed at unity, i.e., a box of size $L$ contains $L^2$ sites. The nearest neighbors of each site are determined by means of the VD construction \cite{okabe_boots_book_00}. The Voronoi cell associated with a lattice site is the area encompassing all points in the system area that are closer to the given lattice site than to any other. Two lattice sites are regarded as nearest neighbors if their Voronoi cells have a common edge. The Delaunay triangulation is the dual of the Voronoi diagram and is defined as the graph of all bonds linking neighboring pairs. The edges of the Voronoi cells are the perpendicular bisectors of the edges in the Delaunay triangulation, and the vertices of the Voronoi diagram are the circumcenters of the triangles in the Delaunay triangulation.

In a random VD lattice, the number of nearest neighbors vary site to site. These connectivity fluctuations introduce topological disorder into the system. Barghathi and Vojta \cite{barghathi_vojta_prl_14} showed that a topological constraint imposed by the Euler equation in a two-dimensional triangulation graph introduces strong anti-correlations between the connectivity fluctuations. This suppresses the disorder effects and results in the modified Harris criterion, $(d+1)\nu>2$, for the stability of a critical point against such topological disorders. More details are discussed in Appendix A.

Our algorithm \cite{puschmann_cain_epjb_15} for computing the Voronoi diagram and Delaunay triangulation for a given set of random points follows a suggestion by Tanemura et al. \cite{tanemura_ogawa_jcp_83}. It relies on the striking ``empty circumcircle property'' of a Delaunay triangulation which states that every triangular facet formed by the nearest-neighbor bonds has an empty circumcircle, i.e., a circumcircle that does not contain any other lattice sites. Our method finds the Voronoi neighbors of a lattice site in two steps: (i) candidates for the neighbors are identified based on their distance. (ii) Using these candidates, we then construct all triangles with empty circumcircles for which the given site is one of the vertices. The resulting algorithm is pretty efficient, finding the Delaunay triangulation of $10^6$ sites takes about 30 seconds on an older Intel core i5 CPU.

In order to study the effects of generic disorder on top of the topological disorder of the random VD lattice, we also consider additional site dilution which is explained in detail in Section \ref{sec6}.

\subsection{Monte Carlo Simulations: Analysis of critical behavior}
\label{sec3-A}

We use a combination of the Metropolis single-spin algorithm \cite{metropolis_rosenbluth_jcp_53} and the Wolff cluster algorithm \cite{wolff_prl_89} to perform our Monte Carlo simulations of the classical Hamiltonian (\ref{eqn.2}). A full Monte Carlo sweep is composed of a Wolff cluster sweep and a Metropolis sweep over the entire lattice. The Wolff cluster flips greatly reduce the critical slowing down, and the Metropolis flips help to equilibrate isolated small clusters of lattice sites, which is particularly useful in the presence of additional site dilution.

To reduce the computational cost in performing these Monte Carlo simulations, we adopt the strategy outlined, e.g., in Refs.\ \cite{vojta_rastko_prb_06,ballesteros_fernandez_prb_98,crewse_vojta_prb_21}. This approach is based on averaging over a large number of disorder configurations while keeping the number of measurement sweeps for each configuration relatively small. This is possible because the Wolff algorithm leads to a short equilibration time. Specifically, we perform simulations over about 1000  to 5000 disorder configurations (depending on system size and disorder strengths) while keeping the number of equilibration sweeps to about 100 to 200 and the number of measurement sweeps to 500. The strategy of simulating many disorder configurations while keeping the number of measurements small reduces the overall variance of various observables. As usual, the quality of the equilibration is checked by comparing simulations with hot and cold starts.

Our primary observable is the order parameter
\begin{equation}
    \mathbf{m}=\frac{1}{N}\sum_{i,\tau}\mathbf{S}_{i,\tau},
\end{equation}
where $N$ is the total number of lattice sites. To identify the critical point and to characterize the phase transition, we lay recourse to the Binder cumulant of the order parameter, defined as
\begin{equation}
    U_{m}=\left[1-\frac{
    \langle|\mathbf{m}|^{4}\rangle}{3\langle|\mathbf{m}|^{2}\rangle^{2}}\right]_{dis}~.
\end{equation}
Here, $\langle \ldots \rangle$ denotes the thermodynamic (Monte Carlo) average while the average over disorder realizations is represented by $\left[ \ldots \right]_{dis}$.
Furthermore, we also compute the order parameter susceptibility given by
\begin{equation}
    \chi= \frac N T \left[\langle|\mathbf{m}|^{2}\rangle-\langle|\mathbf{m}|\rangle^{2}\right]_{dis}~,
\end{equation}
where $T$ is the classical temperature of the system.

Now, we discuss the finite-size scaling methods used to identify and analyze the critical point.
The introduction of quenched randomness breaks the symmetry between the space and imaginary time directions. We thus cannot expect the critical point to be Lorenz-invariant. Instead, the spatial correlation length $\xi$ and the correlation length in imaginary time direction $\xi_{\tau}$ must be treated as independent length scales. Correspondingly, we need to treat the spatial system size $L$ and the size in the imaginary time direction, $L_{\tau}$, as independent parameters. Note that $L_{\tau}$ encodes the inverse physical temperature of the original quantum model (\ref{eqn.1}).
Assuming conventional power-law dynamical scaling $\xi_\tau \sim \xi^z$, the expected finite-size scaling form of the Binder cumulant reads
\begin{equation}
    U_{m}=X_{U}(rL^{1/\nu},L_{\tau}/L^{z}),
    \label{anisotropic}
\end{equation}
where $X_{U}$ is a dimensionless scaling function, $\nu$ is the correlation length critical exponent, $z$ is the dynamical exponent, and $r=(T-T_{c})/T_{c}$ is the reduced distance from criticality.

To find the critical temperature and to measure the value $z$ (which determines the aspect ratio, i.e., the shape of the sample), we follow the anisotropic scaling method described, e.g., in Refs.\ \cite{rieger_young_prl_94, sknepnek_vojta_prl_04}. We use the fact that, at fixed $L$, the Binder cumulant as a function of $L_{\tau}$ has a maximum at imaginary-time size $L_{\tau}^{max}$. The pair of system sizes ($L$, $L_{\tau}^{\rm max}$) gives the shape for which the correlations decay about equally in the space and imaginary time directions; it is called the optimal shape. According to the scaling form (\ref{anisotropic}), the position of the maximum at criticality, $r=0$, is expected to behave as
\begin{equation}
    L_{\tau}^{\rm max} \sim L^{z}.
    \label{eqn.scaling_z}
\end{equation}
Moreover, because the Binder cumulant is a dimensionless quantity, its values $U_{m}^{max}$ at criticality are independent of $L$. Hence, we can use this method to estimate the transition temperature $T_{c}$.

The order parameter and order parameter susceptibility have analogous scaling forms,
\begin{equation}
    m=L^{-\beta/\nu}X_{m}(rL^{1/\nu},L_{\tau}/L^{z})
    \label{eq.scaling1}
\end{equation}
and
\begin{equation}
    \chi=L^{\gamma/\nu}X_{\chi}(rL^{1/\nu},L_{\tau}/L^{z})~,
    \label{eq.scaling2}
\end{equation}
where $\beta$ and $\gamma$ are the order parameter and susceptibility critical exponents. $X_{m}$ and $X_{\chi}$ are dimensionless scaling functions. Once the optimal shapes are found (fixing the second argument of the scaling functions), the finite-size scaling analysis to extract the critical exponent proceeds as usual \cite{cardy_88}.

In our simulations, we use system sizes up to $L=128$ in the space direction and up to $L_\tau=400$ in the imaginary-time direction.

\subsection{Amplitude mode and Wick rotation}
\label{subsec:3.b}

The amplitude mode is an oscillation of the order parameter magnitude in the symmetry-broken (ordered) phase. Since the local degrees of freedom $\mathbf{S}_{i,\tau}$ in the classical Hamiltonian (\ref{eqn.2}) have fixed magnitude, we define a fluctuating local order parameter magnitude $\rho$ via coarse graining. To this end, we average $\mathbf{S}_{i,\tau}$ over
a small cluster consisting of site $i$ and all its (spatial) nearest neighbors,
\begin{equation}
    \rho(\mathbf{x}_{i},\tau)=\frac{1}{\kappa_{i}}\left|\mathbf{S}_{i,\tau}+\sum_{j}\mathbf{S}_{j,\tau}\right|~.
\end{equation}
Here, $\kappa_{i}$ is the number of sites in the local cluster. The fluctuations of this order parameter amplitude can be gleaned from the imaginary-time scalar susceptibility
\begin{equation}
    \chi_{\rho\rho}(\mathbf{x},\tau)=\langle\rho(\mathbf{x},\tau)\rho(0,0)\rangle-\langle\rho(\mathbf{x},\tau)\rangle\langle\rho(0,0)\rangle
    \label{eqn:chirhorho}
\end{equation}
which becomes translationally invariant after disorder averaging. A Fourier transformation then gives the scalar susceptibility $\tilde{\chi}_{\rho\rho}(\mathbf{q},i\omega_{m})$ as a function of wave vector $\mathbf{q}$ and Matsubara frequency $i\omega_m$.

The Monte Carlo simulations yield the imaginary time (or Matsubara frequency) scalar susceptibility. To make contact with the experimentally relevant real-time or real-frequency scalar susceptibility,
we need to perform an analytic continuation (Wick rotation),
\begin{equation}
    \chi_{\rho\rho}(\mathbf{q},\omega)=\tilde{\chi}_{\rho\rho}(\mathbf{q},i\omega_{m}\rightarrow \omega +i0^{+})~.
\end{equation}
 Here, $\omega$ is the real frequency. The (critical part of the) real frequency scalar susceptibility is expected to satisfy the generalized scaling relation \cite{podolsky_sachdev_prb_12,puschmann_crewse_prl_20},
\begin{equation}
    \chi_{\rho\rho}(\mathbf{q},\omega)=\omega^{[(d+z)\nu-2]/(\nu z)}X(\mathbf{q}r^{-\nu},\omega r^{-\nu z})~.
    \label{eqn:scaling}
\end{equation}
Here, $X$ is the scaling function. The Wick rotation of the scalar susceptibility can be implemented as an integral transformation \begin{equation}
    \tilde{\chi}_{\rho\rho}(\mathbf{q},i\omega_{m}) = \frac{1}{\pi}\int_{0}^{\infty} d\omega\chi_{\rho\rho}^{''}(\mathbf{q},\omega)\frac{2\omega}{\omega_{m}^{2}+\omega^2}
    \label{scaling_susceptibility}
\end{equation}
where $\chi_{\rho\rho}^{''}(\mathbf{q},\omega)={\rm Im}\chi_{\rho\rho}(\mathbf{q},\omega)$ is the spectral function.
Unfortunately, this analytic continuation is an ill-posed problem and is sensitive to errors from the numerical data. However, we can overcome this problem by using the maximum entropy method (MaxEnt) \cite{jarrell_guernatis_pr_96}. More details of the technique are given in Appendix B.

Note that we will focus on wave vector $\mathbf{q}=0$ in the following. Nonzero wave vectors lead to extra numerical complications for our random VD lattice because the coordinates of the lattice sites can take arbitrary real values that vary from disorder realization to disorder realization. Thus, the efficient fast-Fourier transformation algorithms cannot be applied directly.

\section{Results}
\label{sec4}

In this section, we present the results for the quantum critical behavior of the superfluid-insulator transition in the Bose-Hubbard model (\ref{eqn.1}) on a random VD lattice. Furthermore, we study the localization properties of the amplitude mode in the symmetry-broken (superfluid) phase in the presence of topological disorder.

\subsection{Analysis of the critical behavior}
\label{subsec4-A}

\begin{figure*}
    \centering
    \includegraphics[width=\textwidth]{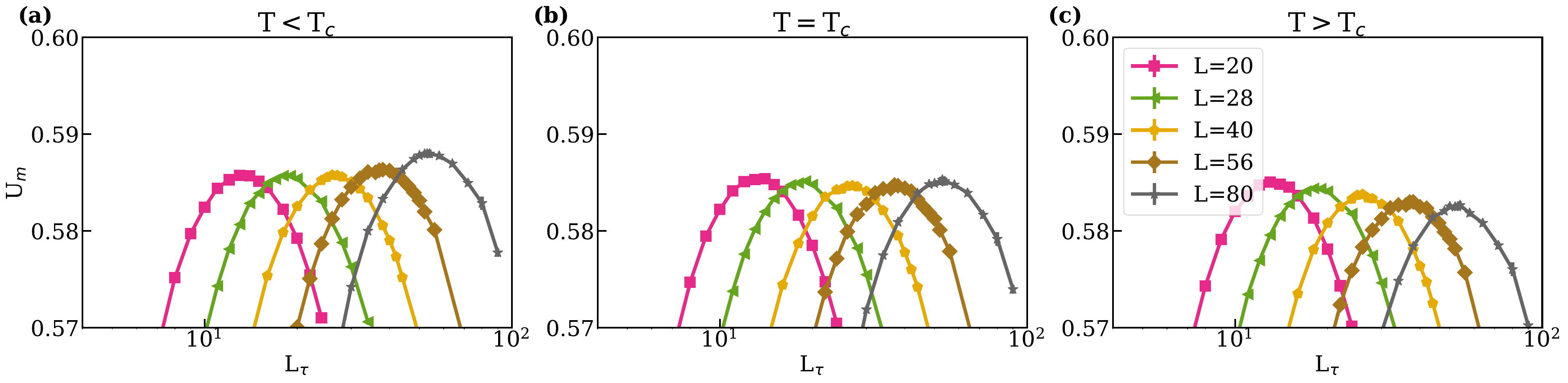}
    \caption{\justifying Binder cumulant $U_m$ as a function of the imaginary-time system size $L_\tau$ for different spatial system sizes $L$ at (a) $T=3.15480<T_{c}$, (b) $T=3.15505=T_{c}$ and (c) $T=3.15530>T_{c}$. The peak values are independent of $L$ at $T_{c}$ whereas they decrease with $L$ for $T>T_c$ and increase with $L$ for $T<T_{c}$. The statistical errors are of the order of the symbol size or smaller.}
    \label{fig:voronoi_anisotropic}
\end{figure*}

As discussed in Sec.\ \ref{sec3-A}, the first step towards analyzing the critical behavior consists in finding the critical point by using the anisotropic finite-size scaling of the Binder cumulant $U_m$. To do so, we compute $U_m$ for a given spatial system size $L$ over a range of imaginary-time system sizes $L_{\tau}$ at different temperatures around the phase transition. We identify the critical temperature of the system as the temperature at which the peak values of the $U_m$ vs.\ $L_\tau$ curves are independent of $L$ (see Fig. \ref{fig:voronoi_anisotropic}).
By this method, the transition temperature of the classical Hamiltonian (\ref{eqn.2}) is found to be $T_{c}=3.15505(25)$.

\begin{figure}
    \centering
    \includegraphics[width=\columnwidth]{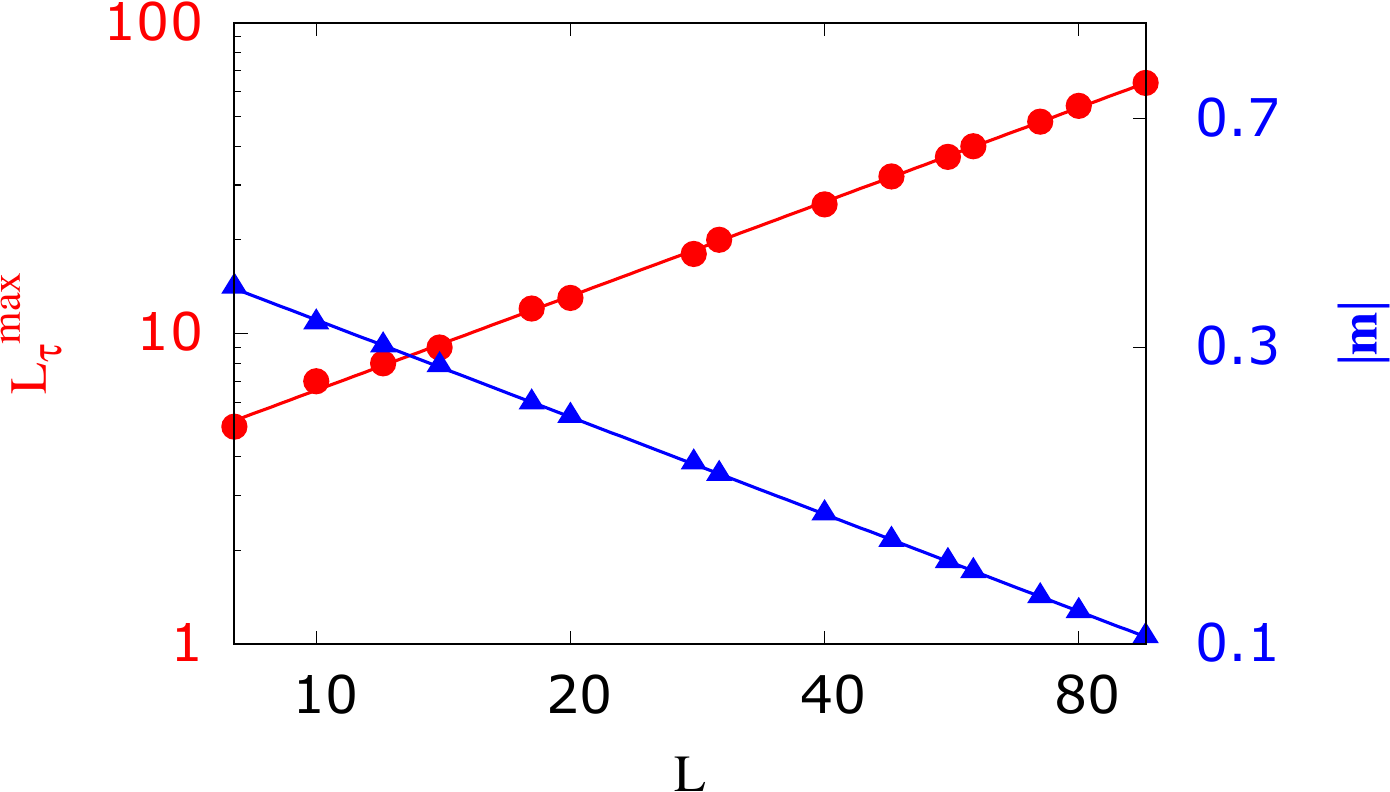}
    \caption{\justifying $L_{\tau}^{\rm max}$ (red circles) and $|\mathbf{m}|$ (blue triangles) as functions of $L$ at $T_{c}$. The lines are fits to the predictions of the scaling relations $L_{\tau}^{\rm max}=aL^{z}$ and $m = aL^{-{\beta}/{\nu}}$. The statistical errors are of the order of the symbol size or smaller.}
    \label{fig:z_and_betabynu}
\end{figure}

\begin{figure}
    \centering
    \includegraphics[width=\columnwidth]{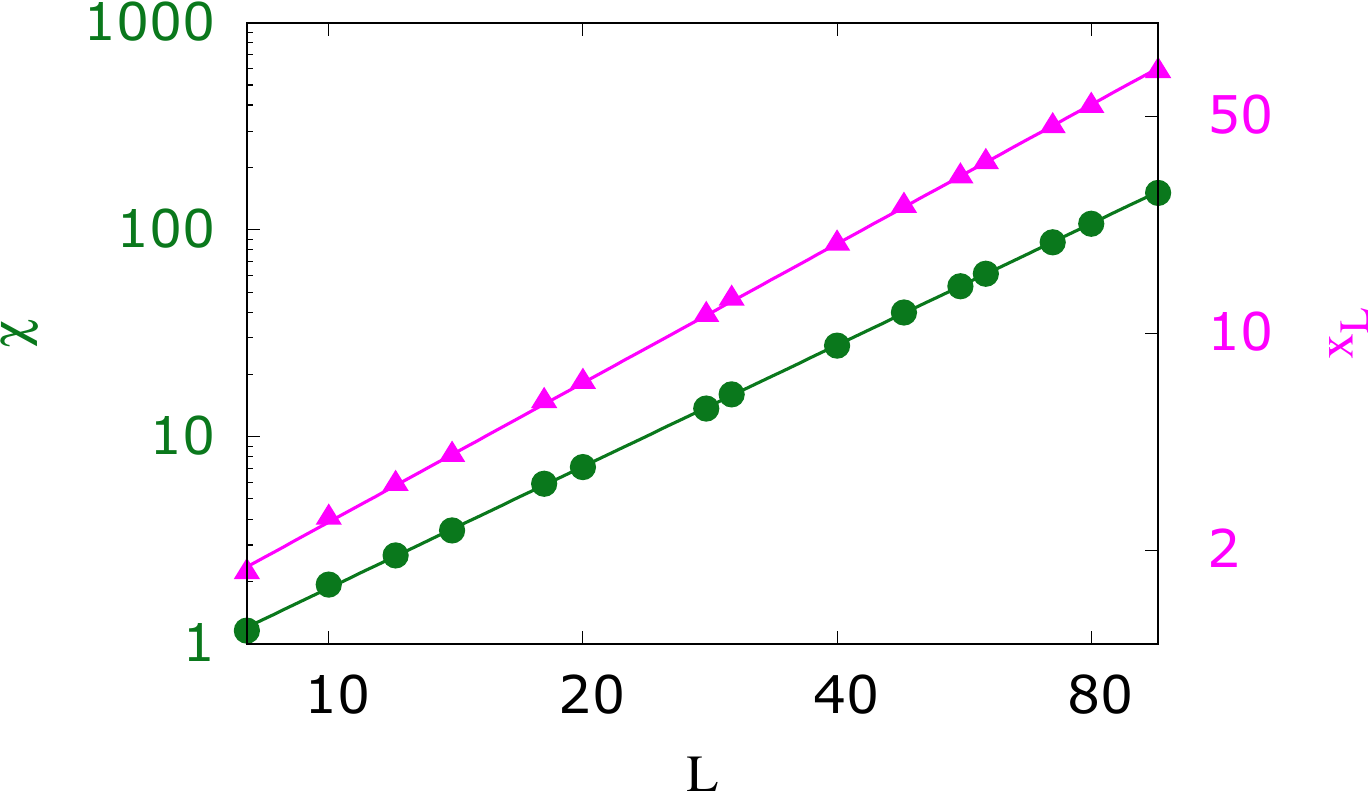}
    \caption{\justifying $\chi$ (green circles) and $x_{L}$ (magenta triangles) as functions of $L$ at $T_{c}$. The lines are fits to the power-law relations $\chi=aL^{{\gamma}/{\nu}}$ and $x_{L}= aL^{{1}/{\nu}}$. The statistical errors are of the order of the symbol size or smaller.}
    \label{fig:nu_and_gammabybynu}
\end{figure}

To obtain the optimal shapes, we determine the position $L_\tau^{\rm max}$ for each $U_m$ vs.\ $L_\tau$ curve (by fitting a quadratic parabola in terms of $\ln{L_{\tau}}$). The dynamical critical exponent $z$ is obtained by fitting the optimal shapes to the relation $L_{\tau}^{\rm max}=aL^{z}$, as shown in Fig.\ \ref{fig:z_and_betabynu}.
Other critical exponents can be measured by analyzing systems of the optimal shapes (fixing the argument $L_\tau/L^z$ in the scaling functions) at or close to $T_c$. The order parameter critical exponent $\beta/\nu$ is obtained from the decay of the order parameter with $L$ right at criticality, $m = aL^{-{\beta}/{\nu}}$.  Similarly, the critical exponent $\gamma/\nu$ of the order parameter susceptibility is obtained from $\chi = aL^{{\gamma}/{\nu}}$ at criticality, see Fig.\ \ref{fig:nu_and_gammabybynu}.
To get the correlation length critical exponent $\nu$, we first define the observable, $x_{L}=|(d/dT)\ln{|\textbf{m}|}|$. The corresponding scaling form can be obtained by differentiating the scaling form (\ref{eq.scaling1}) of the order parameter with respect to $T$. The resulting $L$ dependence at criticality is given by $x_{L} = aL^{{1}/{\nu}}$.

The values of critical exponents resulting from this analysis are tabulated in Table \ref{table:1} along with the exponents obtained for the clean three-dimensional XY universality class \cite{compostrini_hasenbusch_prb_06} and those for the superfluid-Mott glass transition with generic disorder \cite{vojta_crewse_prb_16}.
\begin{table}
\renewcommand*{\arraystretch}{1.2}
\begin{tabular*}{\columnwidth}{c @{\extracolsep{\fill}} c c c}
 \hline\hline
 Exponent & Clean \cite{compostrini_hasenbusch_prb_06} & Generic disorder \cite{vojta_crewse_prb_16} & VD lattice \\
 \hline
$\nu$ & 0.6717 & 1.16(5) & 0.672(8) \\
${\beta}/{\nu}$ & 0.519 & 0.48(2) &  0.520(4) \\
${\gamma}/{\nu}$ & 1.962 & 2.52(4) & 1.950(10) \\
$z$ & 1 & 1.52(3) & 1.008(9)\\
 \hline\hline
\end{tabular*}
\caption{\justifying Critical exponents of the superfluid-insulator quantum phase transition on the random VD lattice, compared to the clean case \cite{compostrini_hasenbusch_prb_06} and the case of generic disorder \cite{vojta_crewse_prb_16}. The quoted error bars include the error due to the uncertainty of $T_c$ as well as the robustness of the fits against removing data points at the ends of the $L$ range.}
\label{table:1}
\end{table}
The table clearly shows that the critical exponents of the superfluid transition on the random VD lattice agree with those of the equivalent translationally invariant (clean) system rather than with the exponents of superfluid transition in the presence of generic disorder. This is consistent with the modified Harris criterion derived by Barghathi and Vojta \cite{barghathi_vojta_prl_14},  which states that a clean critical point is stable against the topological disorder of a random VD lattice if $(d+1)\nu > 2$. We note that similar behavior was observed at the superconducting transition in quasi-crystals \cite{Eric_quasi}: It was shown that even though the local connectivity in the quasi-crystal varies from site to site, the critical behavior associated with the transition (within a BCS approximation) was in line with that of the clean system. Finally, the effects of additional generic uncorrelated disorder on top of the topological disorder of the random VD lattice will be discussed in Section \ref{sec6}.

\subsection{Analysis of the amplitude mode}
\label{subsec:4.2}

We now discuss the behavior of the amplitude (Higgs) mode close to the quantum phase transition. It is known that, in the absence of disorder, the system features a well-defined, soft-gapped amplitude mode characterized by a peak in the spectral function $\chi_{\rho\rho}^{''}(\mathbf{q}=0,\omega)$ at the Higgs energy $\omega_{H}$ \cite{gazit_podolsky_prl_13,crewse_vojta_prb_21,puschmann_crewse_prl_20}. This peak survives as the quantum phase transition is approached from the superfluid side. In agreement with the scaling form (\ref{eqn:scaling}), the Higgs energy vanishes as $\omega_{H} \sim r ^{z\nu}$. In contrast, a Higgs peak is not observable in the presence of generic disorder where the scalar response is broad and non-critical \cite{puschmann_crewse_prl_20, crewse_vojta_prb_21,puschmann_getelina_annals_21}, indicating that the amplitude mode is localized.

\begin{figure}
    \centering
    \includegraphics[width=\columnwidth]{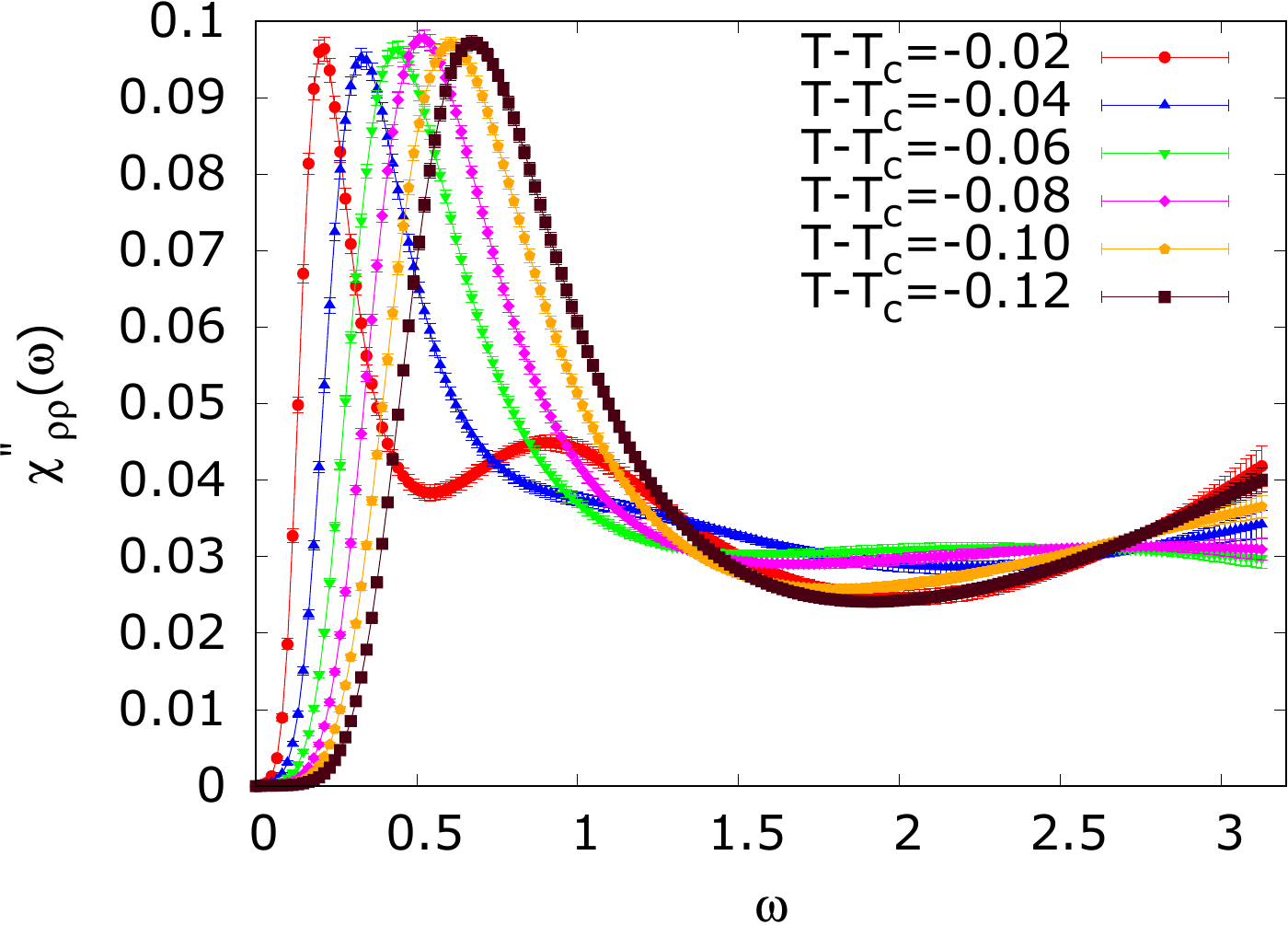}
    \caption{\justifying Spectral function $\chi^{''}_{\rho\rho}(\mathbf{q}=0,\omega)$ as a function of real frequency $\omega$ for different distances to the (classical) critical temperature in the superfluid phase. The simulation is performed for a lattice with $L=L_{\tau}=128$ (reflecting the dynamical exponent $z=1$).}
    \label{fig:voronoi_peaks_without_dilution}
\end{figure}

\begin{figure}
    \centering
    \includegraphics[width=\columnwidth]{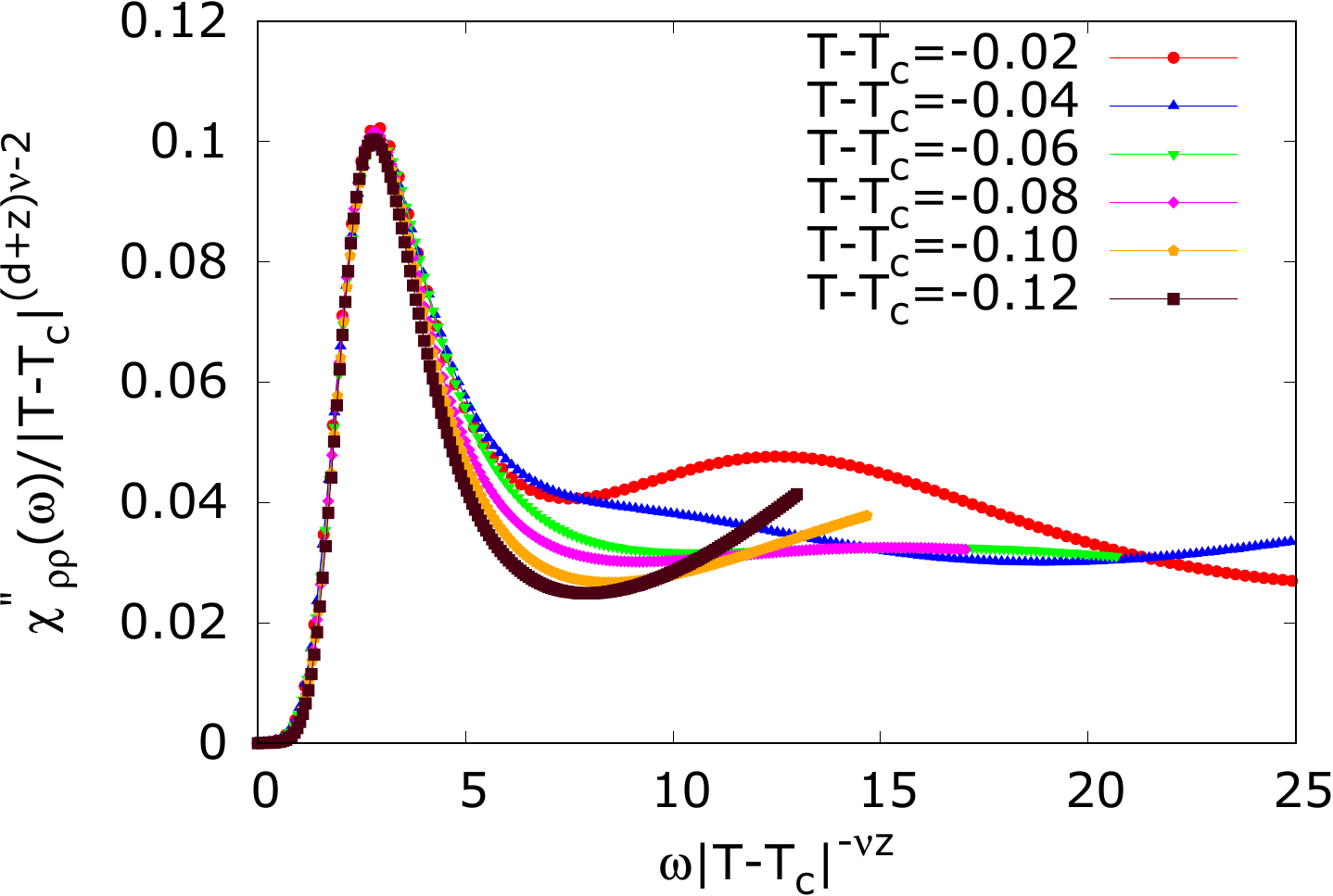}
    \caption{\justifying Scaling plot of the spectral density $\chi^{''}_{\rho\rho}(\mathbf{q}=0,\omega)$ on the superfluid side of the quantum phase transition for the Voronoi lattice, as suggested by the scaling form (\ref{eqn:scaling}). The exponent values, $\nu=0.6717$ and $z=1$ with $d=2$, belong to the clean XY universality class.}
    \label{fig:voronoi_scaling_collapse}
\end{figure}

In Fig.\ \ref{fig:voronoi_peaks_without_dilution}, we present the scalar spectral function $\chi_{\rho\rho}^{''}(\mathbf{q}=0,\omega)$ for the Hamiltonian (\ref{eqn.2}) defined on a random VD lattice
at different distances from criticality in the superfluid phase.
The data clearly demonstrate that the amplitude mode in this topologically disordered system behaves analogously to the clean case, as  $\chi_{\rho\rho}^{''}(\mathbf{q}=0,\omega)$ features a well-pronounced low-energy peak that softens as the critical point is approached.  No indications of spatial localization are observed. Additionally, the Higgs peak in the scalar spectral function shows an excellent scaling collapse according to Eq.\ (\ref{eqn:scaling}) using the clean critical exponents, as seen in Fig.\ \ref{fig:voronoi_scaling_collapse}.
The results for this topologically disordered system are thus qualitatively different from those obtained on a square lattice with site dilution \cite{crewse_vojta_prb_21,puschmann_crewse_prl_20} which represents the generic disorder case.

Our findings give us important information on the physical origins of the localization mechanism of the amplitude mode. On the one hand,  a tight-binding model of non-interacting particles on a random VD lattice has been shown to feature Anderson localization \cite{puschmann_cain_epjb_15}. In other words, the connectivity disorder anticorrelations of the random VD lattice do not qualitatively affect the Anderson localization mechanism.  On the other hand, we found here that the amplitude mode close to the superfluid-insulator transition on a random VD lattice does not become localized, in contrast to the generic disorder case. This indicates that the localization of the amplitude mode observed in Refs.\ \cite{crewse_vojta_prb_21,puschmann_crewse_prl_20} may not be governed by an Anderson-type localization mechanism. The modified Harris criterion \cite{barghathi_vojta_prl_14} states that the topological disorder of the random VD lattice is an irrelevant operator at the critical point of a (2+1) dimensional XY model. Our results are therefore consistent with the localization of the amplitude mode (close to the phase transition) being tied to the critical behavior.

\begin{figure}
    \centering
    \includegraphics[width=\columnwidth]{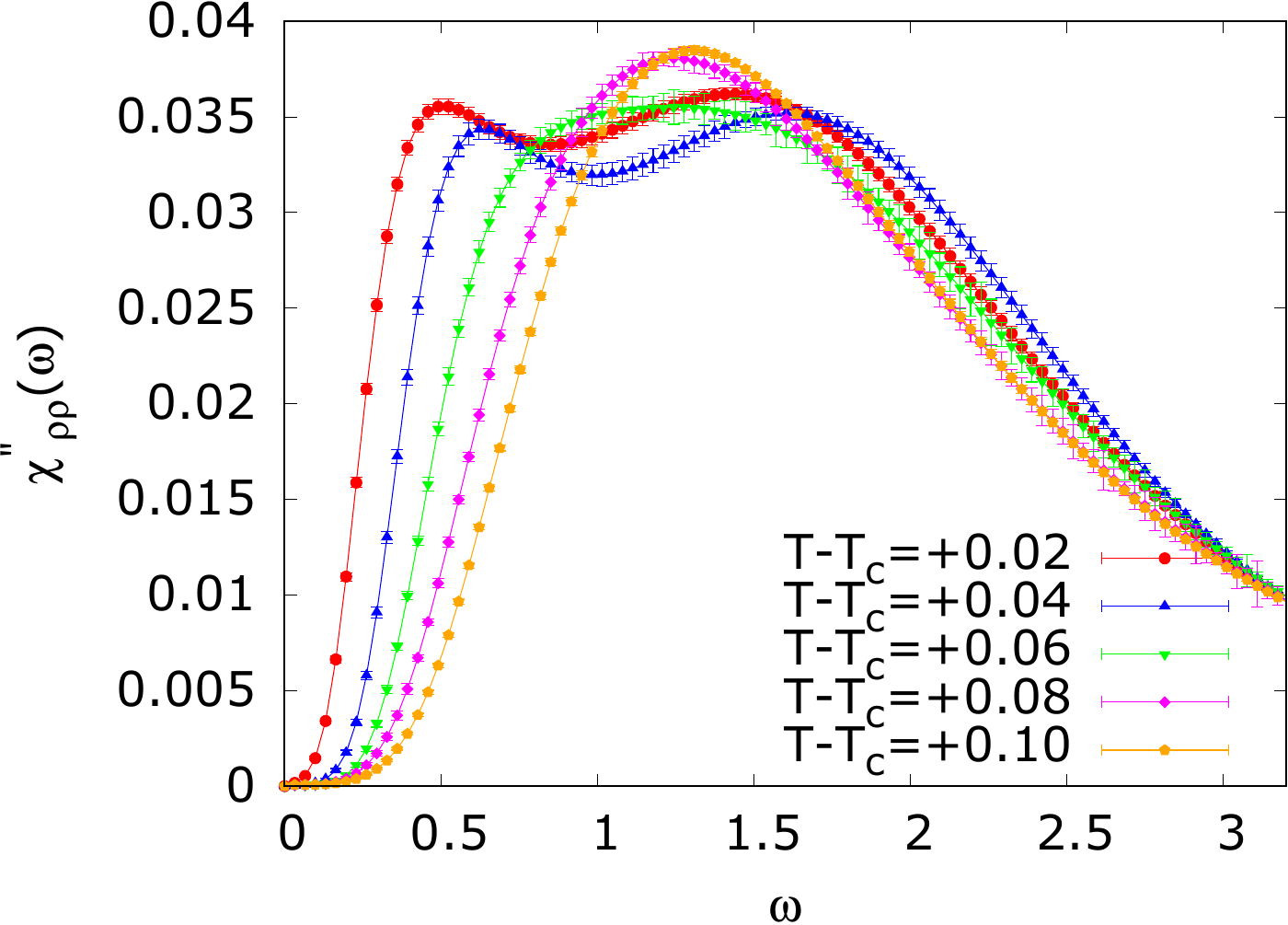}
    \caption{\justifying Spectral function $\chi^{''}_{\rho\rho}(\mathbf{q}=0,\omega)$ as a function of real frequency $\omega$ for different distances to the critical temperature in the Mott phase. The simulations are performed for a lattice with $L=L_{\tau}=128$.}
    \label{fig:voronoi_peaks_without_dilution_mottphase}
\end{figure}

What about the scalar susceptibility in the Mott phase (the disordered phase)? Since the order parameter vanishes in the thermodynamic limit, one might not expect to observe a Higgs resonance in this case. However, simulations of a clean Bose-Hubbard model \cite{chen_liu_prl_13} identified a spectral peak associated with the Higgs mode not just in the superfluid phase but also in the insulating phase sufficiently close to the critical point, such that a local order parameter can be defined on a large but finite length scale.  Analogously, the scalar spectral density of our topologically disordered system features peaks in the Mott phase
(i.e., for classical temperatures slightly above $T_c$), as shown in Fig.~\ref{fig:voronoi_peaks_without_dilution_mottphase}. These peaks soften on approaching the critical point, but they are less sharp than the Higgs peaks in the superfluid phase. Furthermore, just as in Ref.\ \cite{chen_liu_prl_13}, the spectral density close to criticality shows a double peak structure emanating from the interplay of the amplitude mode with the generic critical order parameter fluctuations.
\section{Inhomogeneous mean-field theory}
\label{sec5}

To gain further insight into the behavior of the collective excitations of the bosonic Hubbard model (\ref{eqn.1}) on a random VD lattice, we apply an inhomogeneous mean-field theory \cite{puschmann_crewse_prl_20, puschmann_getelina_annals_21}.
 The excitations are obtained from expanding the Hamiltonian about the spatially inhomogeneous mean-field solution.
At the Gaussian level, this results in noninteracting bosonic excitations. This theory generalizes the approach of Refs.\ \cite{AltmanAuerbach02,Pekkeretal12} to the disordered
case and is related to the bond-operator method for Heisenberg magnets \cite{VojtaM13}. In the presence of disorder, the approach captures Anderson localization physics but not
the mode-mode coupling and renormalizations near criticality, allowing us to disentangle the mechanisms for the amplitude mode localization or lack thereof.

To derive the mean-field theory, we truncate the local Hilbert space at lattice site $j$ to three basis states, $|-_j\rangle,  |0_j\rangle$, and $|+_j\rangle$, representing
the particle numbers $n_j = \bar n -1, \bar n$, and $\bar n +1$, respectively. The variational ground state wave function is written in product form,
$|\Phi_0 \rangle = \prod_j |\phi_{0j}\rangle$ with
\begin{eqnarray}
|\phi_{0j}\rangle &=& \cos(\theta_j/ 2) |0_j\rangle  \nonumber \\
                   && + \sin(\theta_j/ 2) \left( e^{i \eta_j} |+_j\rangle + e^{-i \eta_j} |-_j\rangle \right)/\sqrt 2~.
\end{eqnarray}
The parameters (mixing angles) $\theta_j$ control the character of the ground state, yielding a Mott insulator for $\theta_j=0$ and a superfluid for $\theta_j>0$.
The parameter $\eta_j$ represents the phase of the local superfluid order parameter $\langle a_j^\dagger \rangle \propto  \psi_j = \sin \theta_j e^{-i \eta_j}$.

Minimizing the ground state energy $E_0 = \langle \Phi_0 | H | \Phi_0 \rangle$ w.r.t. to the mixing angles $\theta_j$ and phases $\eta_j$ gives
uniform $\eta_j=\eta=\textrm{const}$ (which we set to zero in the following) while the mixing angles fulfill the mean-field
equations
\begin{equation}
U_i \sin \theta_i = 4 \bar n \cos \theta_i \sum_j J_{ij} \sin \theta_j~.
\label{eq:MF}
\end{equation}
In the presence of randomness, this large set of coupled nonlinear equations needs to be solved numerically  \cite{puschmann_crewse_prl_20, puschmann_getelina_annals_21}.

\begin{figure}
    \centering
    \includegraphics[width=\columnwidth]{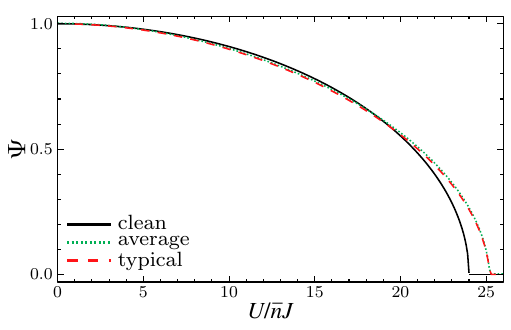}
    \caption{\justifying Superfluid order parameter $\psi$ of the Bose-Hubbard model (\ref{eqn.1}) on a random VD lattice vs.\ $U/(\bar n J)$, computed from the mean-field equations (\ref{eq:MF}). Both the average local order parameter and the typical value (geometric average) are shown. For comparison, the corresponding curve for a clean, triangular lattice is also shown. The data are averages over 1000 disorder configurations for a system size for $L=128$. The resulting statistical errors are below the line thickness. }
    \label{fig:mf-op}
\end{figure}

Figure \ref{fig:mf-op} presents the resulting order parameter $\psi$ of the Bose-Hubbard model (\ref{eqn.1}) on a random VD lattice as a function of $U/J$.
The data shows that the superfluid-insulator transition happens at $U/(\bar n J) \approx 25.3$. This is close to the value for the regular (clean) triangular lattice, $U/(\bar n J) = 24$ which is included in the figure for comparison. (This makes sense because the average coordination number of a random VD lattice is exactly 6, identical to the coordination number of the triangular lattice.) The figure also shows that the average order parameter $(1/N)\sum_j \psi_j$ and the typical value $\exp[(1/N)\sum_j \ln \psi_j]$ are almost indistinguishable. This indicates that the order parameter fluctuations are weak, much smaller than those of a site-diluted Bose-Hubbard model at moderate dilutions (see Fig.\ 2 of Ref.\ \cite{puschmann_getelina_annals_21}).

In order to investigate excitations of the mean-field ground state, we now transform the basis in the Hilbert space associated with site $j$
from $|-_j\rangle,  |0_j\rangle$, $|+_j\rangle$ to a new orthonormal  basis consisting of $|\phi_{0j}\rangle, |\phi_{Hj}\rangle, |\phi_{Gj}\rangle$.
The states
\begin{eqnarray}
|\phi_{Hj}\rangle &=& \sin(\theta_j/ 2) |0_j\rangle - \cos (\theta_j / 2) \left(  |+_j\rangle + |-_j\rangle \right)/\sqrt{2} \quad \nonumber \\
|\phi_{Gj}\rangle &=&  i \left(  |+_j\rangle - |-_j\rangle \right)/\sqrt{2}
\end{eqnarray}
correspond to fluctuations of the order parameter amplitude and phase, respectively, w.r.t.\ to $|\phi_{0j}\rangle$.
We now expand the Hamiltonian (\ref{eqn.1}) to quadratic order in the boson operators $b^\dagger_{Hj}$ and $b^\dagger_{Gj}$ that create these states out of the fictitious vacuum (ground state). The resulting fluctuation Hamiltonian decouples into amplitude and phase parts, $H_{MF}= E_0 + H_H + H_G$, which both take the form
\begin{eqnarray}
H_\alpha= \sum_i A_{\alpha i} b^\dagger_{\alpha i} b_{\alpha i} + \sum_{\langle ij \rangle} B_{\alpha ij} (b^\dagger_{\alpha i} + b_{\alpha i})(b^\dagger_{\alpha j} + b_{\alpha j})~,~
\label{eq:H_H}
\end{eqnarray}
($\alpha=H,G$). The coefficients $A_{\alpha i}$ and $B_{\alpha ij}$ are functions of the mixing angles $\theta_j$.
$H_H$ and $H_G$ can be diagonalized independently by bosonic Bogoliubov transformations,
$b_{\alpha j} = \sum_k ( u_{\alpha jk} d_{\alpha k} + v_{\alpha jk}^\ast d_{\alpha k}^\dagger )$. These Bogoliubov transformations have to be performed numerically because of the randomness.
The importance of higher-order (non-Gaussian) fluctuations not captured in $H_{H}$ and $H_{G}$ is governed by the Ginzburg criterion \cite{ginzburg_spss_61}. As the bare correlation length of the lattice model (Eq. \ref{eqn.1}) is of the order of the lattice constant, the width of the asymptotic critical region is expected to be sizable ($r=(U-U_c)/U_c$ of order unity). This implies that Gaussian fluctuations are not expected to capture the complete physics of the amplitude and phase fluctuations in the parameter region of interest.

\begin{figure}
    \centering
    \includegraphics[width=\columnwidth]{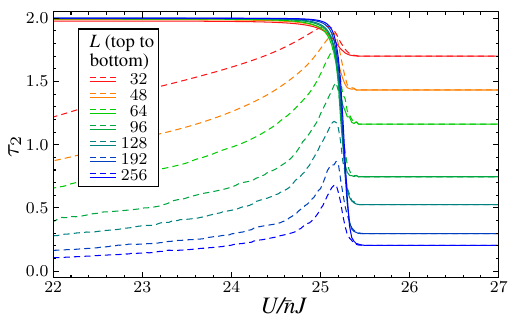}
    \caption{\justifying Generalized dimension $\tau_2(0)$ of the lowest-energy Goldstone (solid lines) and Higgs modes (dashed lines), in Gaussian approximation, vs.\ interaction $U/(\bar n J)$ for different system sizes $L$.  The data are averages over 1000 disorder configurations for a system size for $L=128$. }
    \label{fig:tau2}
\end{figure}

The localization properties of the eigenstates of $H_H$ and $H_G$ can be characterized by their inverse participation numbers. Following Ref.\
\cite{VojtaM13}, the inverse participation number of eigenstate $k$ is given by
 $P^{-1}(k)=\sum_j (|u_{\alpha j0}|^2 - |v_{\alpha j0}|^2)^2$,
and the corresponding generalized dimension reads $\tau_2(k) =   \ln P(k) / \ln L$ (for details see Ref.\ \cite{puschmann_getelina_annals_21}).
Figure \ref{fig:tau2} shows the dimension $\tau_2(0)$ of the lowest-energy Goldstone and Higgs excitations of the Bose-Hubbard model (\ref{eqn.1}) on a random VD lattice as a function of $U/J$.
In the insulating phase, $U/(\bar n J) \gtrsim 25.3$, both excitations are degenerate and strongly localized because $\tau_2$ rapidly decreases towards zero with increasing system size. In the superfluid phase, $U/(\bar n J) \lesssim 25.3$, in contrast, the two excitations behave differently. The lowest Higgs mode remains strongly localized, whereas the lowest Goldstone mode rapidly delocalizes; its $\tau_2$ increases with increasing $L$ and approaches the embedding dimension $d=2$. The delocalization of the Goldstone mode agrees with a general symmetry analysis \cite{GurarieChalker02,GurarieChalker03} and with explicit results for Goldstone modes in a number of systems.
We note that the delocalization of the Goldstone mode within the Bogoliubov Hamiltonian $H_G$ does \emph{not} contradict the results of Ref \cite{puschmann_cain_epjb_15}  which found that noninteracting particles on a random Voronoi-Delaunay lattice Anderson-localize. As is emphasized in Ref. \cite{GurarieChalker03}, the matrix elements in an effective Hamiltonian for bosonic excitations are not independent because they have to fulfill constraints that guarantee the positivity of the spectrum. In $H_G$, correlations between the matrix elements arise due to the dependence of the coefficients $A$ and $B$ on the mixing angles $\theta_j$. It is well known that disorder correlations can modify the localization properties of the eigenstates.

Interestingly, the behavior of the amplitude mode for the Bose-Hubbard model on a random VD lattice agrees with that of the site-diluted Bose-Hubbard model \cite{puschmann_crewse_prl_20, puschmann_getelina_annals_21}. In both cases, the mean-field theory predicts that the lowest Higgs excitation is spatially localized. However, for the random VD lattice, this does not agree with the results of the Monte Carlo simulations reported earlier in Sec.\ \ref{subsec:4.2} where it was found that the amplitude mode on the VD lattice behaves just as in the clean case.
As the inhomogeneous mean-field approach captures the Anderson localization physics but not the mode-mode coupling and renormalization of the full many-particle problem, the discrepancy between the Monte Carlo results and the mean-field results for the amplitude mode on the VD lattice provides further evidence for the amplitude mode localization or lack thereof not being driven by an Anderson localization mechanism.

\section{Adding additional uncorrelated disorder}
\label{sec6}

\subsection{Bose-Hubbard model on a site-diluted random VD lattice}
\label{subsec6.1}

The Monte Carlo results reported in Sec.~\ref{sec4} demonstrated that the amplitude mode remains delocalized in the presence of the topological disorder of a random VD lattice. The results also suggested that this stems from the fact that the topological disorder is an irrelevant perturbation at the clean superfluid-insulator critical point, and the transition belongs to the same universality class as the translationally invariant model. To provide further evidence for this hypothesis, we now add generic uncorrelated disorder in the form of site dilution to the topological disorder of the random VD lattice. This uncorrelated disorder is a relevant perturbation at the clean superfluid-insulator critical point. If the hypothesis is correct, we thus expect not only a crossover to the disordered critical behavior of Ref.\ \cite{vojta_crewse_prb_16}, we also expect the amplitude mode to become localized.

After introducing random vacancies into the Bose-Hubbard Hamiltonian (\ref{eqn.1}) on a random VD lattice, we carry out the quantum-to-classical mapping and arrive at a classical XY model on a layered VD lattice described by the Hamiltonian
\begin{equation}
    H_{\rm dis}=-\sum_{\left<i,j\right>,\tau}J^{s}_{ij}\epsilon_{i}\epsilon_{j}\mathbf{S}_{i,\tau}\cdot\mathbf{S}_{j,\tau}-J^{\tau}\sum_{i,\tau}\epsilon_{i}\mathbf{S}_{i,\tau}\cdot\mathbf{S}_{i,\tau+1}~.
    \label{eqn.2b}
\end{equation}
Here, the independent quenched random variables $\epsilon_i$  take values 0 and 1 with probabilities $p$ and $1-p$, respectively. Since the positions of the vacancies do not depend on the imaginary-time coordinate $\tau$, the disorder is perfectly correlated in the imaginary-time direction (columnar disorder perpendicular to the VD layers).

\subsection{Critical Behavior}
\label{sec6.2}

We perform Monte Carlo simulations of the site-diluted XY model (\ref{eqn.2b}) for six different vacancy concentrations, namely $p= 0.125$, $0.200$, $0.270$  $0.300$, $0.400$ and $0.415$. We analyze the data employing anisotropic finite-size scaling as outlined in Sec.\ \ref{sec3-A}.
The resulting critical temperatures are listed in Tab.~\ref{table:2}.
\begin{table}
\renewcommand*{\arraystretch}{1.2}
\begin{tabular*}{\columnwidth}{c @{\extracolsep{\fill}}  c}
 \hline\hline
 Site dilution $p$ & Critical temperature $T_{c}$  \\
 \hline
0.000   & 3.15505(25) \\
0.125 & 2.850(2) \\
0.200 & 2.648(2) \\
0.270 & 2.448(2) \\
0.300 & 2.356(4) \\
0.400 & 2.020(4) \\
0.415 & 1.960(4) \\
\hline\hline
\end{tabular*}
\caption{\justifying Transition temperatures of the  XY model (\ref{eqn.2b}) with correlated site dilution. }
\label{table:2}
\end{table}

\begin{figure}
    \centering
    \includegraphics[width=\columnwidth]{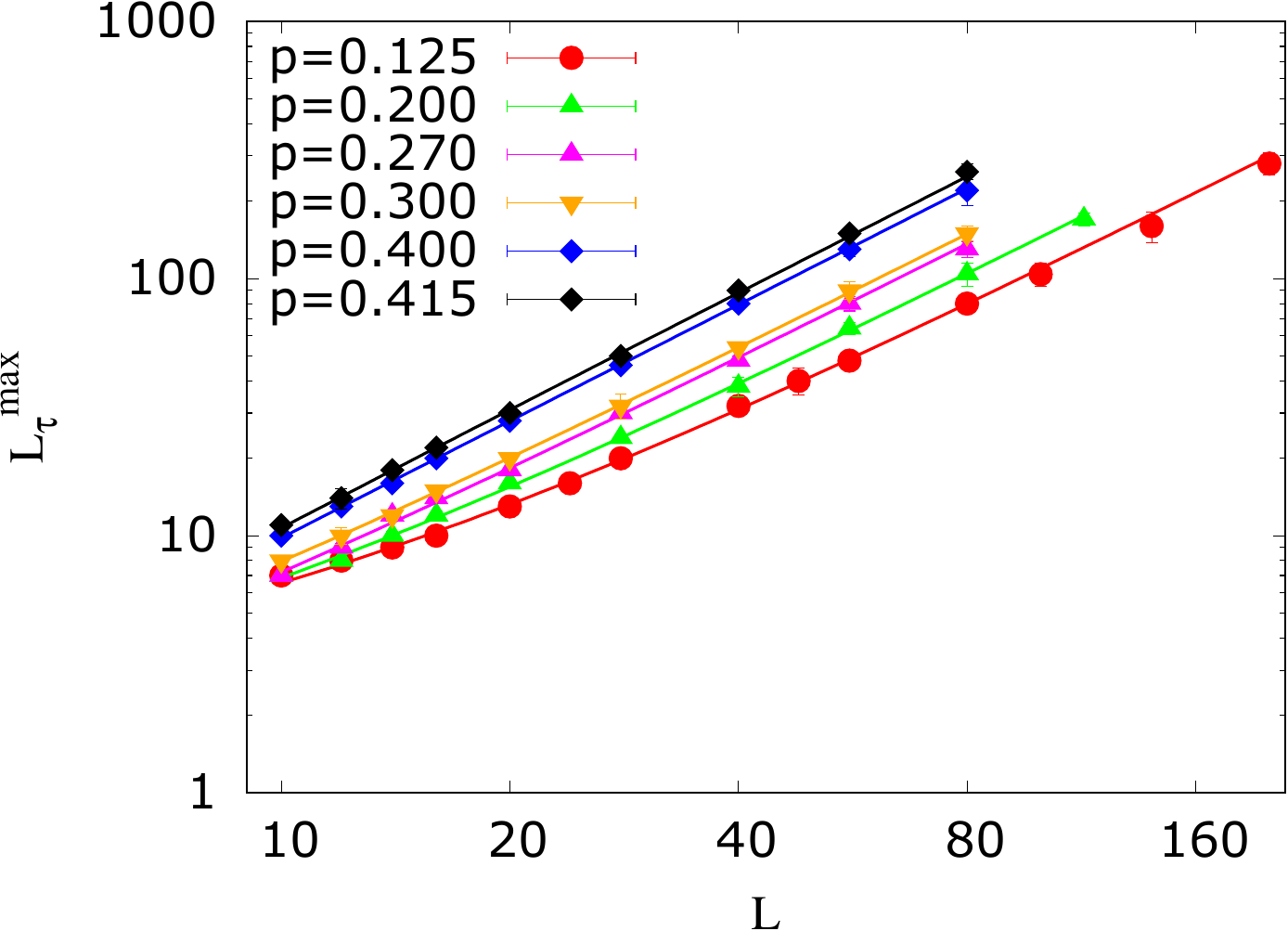}
    \caption{\justifying Double logarithmic plot of $L_{\tau}^{\rm max}$ vs.\ $L$ at $T_{c}$ for various vacancy concentrations $p$. The solid lines represent a combined fit to $L_{\tau}^{\rm max}=aL^{z}(1+bL^{-\omega})$ with universal $z$ and $\omega$ but dilution-dependent coefficients $a$ and $b$. The fit gives $z=1.50(3)$ and $\omega=1.2(2)$. }
    \label{fig:z_with_correction}
\end{figure}

As explained in Sec.~\ref{sec3-A}, the dynamical exponent $z$ can be gleaned from the dependence on the spatial system size $L$ of the peak position $L_{\tau}^{\rm max}$ of the Binder cumulant curves right at the critical temperature. The corresponding data for all dilution values are presented in Fig.\ \ref{fig:z_with_correction}.
In contrast to the corresponding data for the undiluted random VD lattice (Fig.\ \ref{fig:z_and_betabynu}), the data in Fig.~\ref{fig:z_with_correction} display significant deviations from pure power-law behavior $L_{\tau}^{\rm max} \sim L^{z}$, in particular for lower dilutions $p$. This deviation can be attributed to a cross-over from the clean critical behavior to the disordered critical behavior
in response to the (uncorrelated) random-mass disorder generated by the vacancies. To account for this crossover, we include a correction-to-scaling term via $L_{\tau}^{\rm max}=aL^{z}(1+bL^{-\omega})$. The correction term is governed by the irrelevant exponent $\omega$.

The exponents $z$ and $\omega$ are assumed to be universal (independent of the dilution $p$) whereas the coefficients $a$ and $b$ depend on $p$.
Consequently, we perform a combined fit that includes the data for all six dilutions, with universal $z$ and $\omega$ but nonuniversal $a$ and $b$; it yields $z=1.50(3)$ and $\omega=1.2(2)$. The fit is of good quality, giving a reduced $\tilde{\chi}^2 \approx 0.53$. The leading corrections to scaling seem to vanish around $p=0.400$, where the coefficient $b$ of the correction term changes sign. This implies that for $p \approx 0.400$, a simple power-law fit should suffice.  Performing power-law fits for $p=0.400$ and $p=0.415$ yields $z=1.495$ and $z=1.529$, respectively. These values agree with the estimate provided by the combined fit within the statistical errors.

Comparing our results for the site-diluted random VD lattice to those for the site-diluted square lattice \cite{vojta_crewse_prb_16} (see also Tab.\ \ref{table:1}), we notice that values of the dynamical exponent $z$ agree very well, whereas the values of the irrelevant exponent $\omega$ differ from each other. This dichotomy can be explained by the fact that the value of the exponent $z$ reflects the asymptotic critical behavior governed by the finite-disorder renormalization group fixed point. In contrast, the value of $\omega$ is related to the leading irrelevant operator at this fixed point which can be different for different models. Specifically, the Bose-Hubbard model on the diluted random VD lattice contains additional (anti-correlated) topological disorder which may affect the crossover scaling and thus the value of $\omega$.

\begin{figure}
    \centering
    \includegraphics[width=\columnwidth]{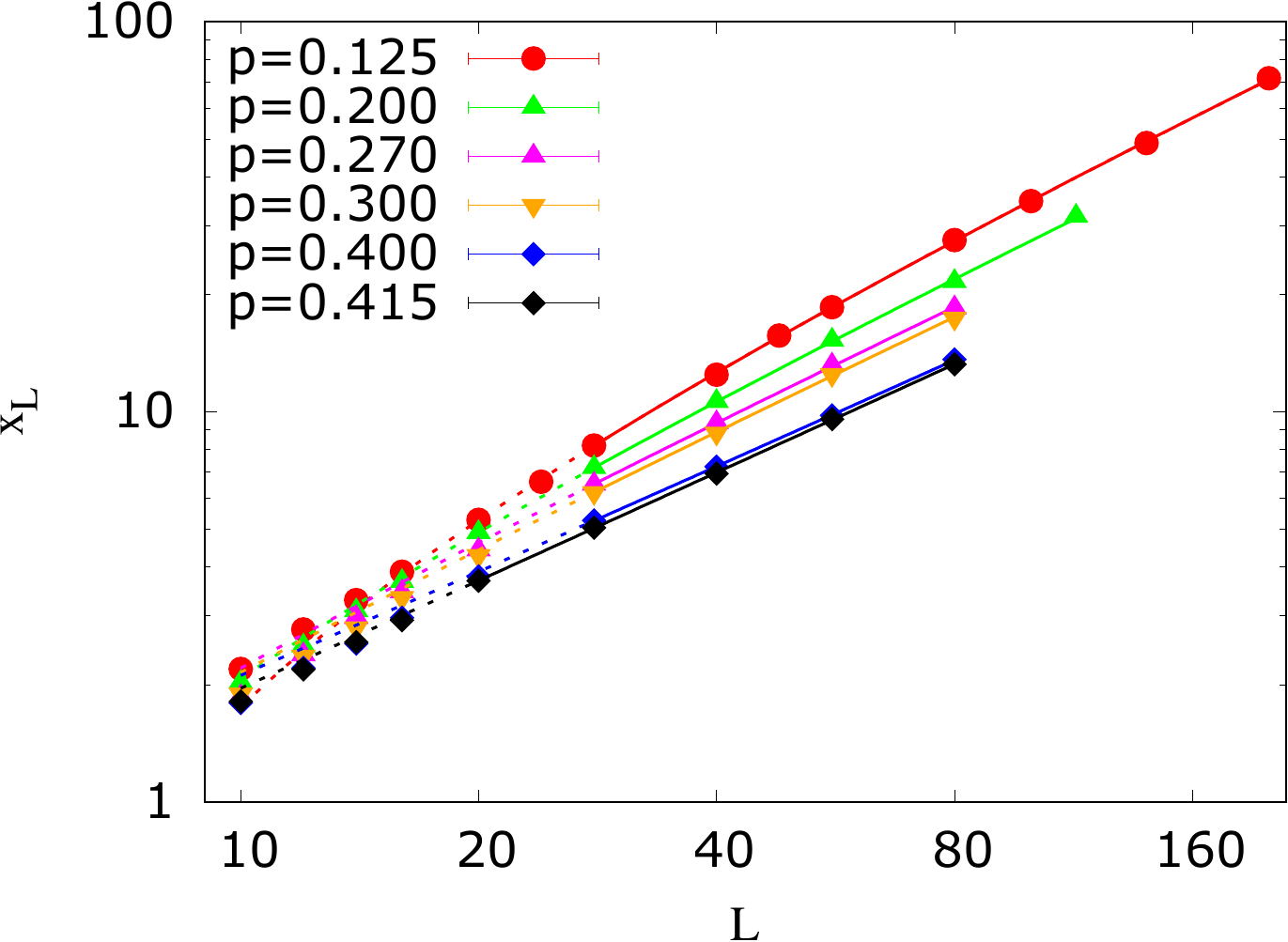}
    \caption{\justifying Double logarithmic plot of $x_{L}=|d\ln{|\textbf{m}|}/dT|$  as function of $L$ near $T_{c}$ for various dilutions $p$.
    The solid lines represent a combined fit of the data for all dilutions with $x_{L}=aL^{1/\nu}(1+bL^{-\omega})$ with universal $\nu$ and $\omega$ but nonuniversal coefficients $a$ and $b$. It gives $\nu=1.08(2)$ and $\omega=0.52(5)$. The dotted lines mark smaller system sizes that are not included in the fit.}
    \label{fig:onebynu_with_correction}
\end{figure}

Having found the optimal sample shapes, we can now determine the exponents $\beta/\nu$, $\gamma/\nu$, and $\nu$ by analyzing the system size dependence of the order parameter, the order parameter susceptibility, and  $x_{L}=|(d/dT)\ln{|\textbf{m}|}|$ at the critical temperature, just as in Sec.\ \ref{subsec4-A}. To capture deviations from pure power-law behavior, we include corrections to scaling terms, and fit the data with the functional forms $m=aL^{-\beta/\nu}(1+bL^{-\omega})$, $\chi=aL^{\gamma/\nu}(1+bL^{-\omega})$, and $x_{L}=aL^{1/\nu}(1+bL^{-\omega})$. For each quantity,
we perform a combined fit of the curves for all dilutions with universal exponent values but dilution-dependent coefficients $a$ and $b$. (In some cases, we need to exclude smaller system sizes to improve the quality of fit and achieve an acceptable $\bar{\chi}^2$.)
As an example, the analysis of $x_L$ leading to a value for the exponent $\nu$ is shown in Fig.~\ref{fig:onebynu_with_correction}.
These fits give the exponent values  $\beta/\nu=0.50(3)$,  $\gamma/\nu=2.47(4)$, and $\nu=1.08(2)$. They all agree within their error bars with the exponents reported for the site-diluted square lattice system \cite{vojta_crewse_prb_16} (see also Tab.\ \ref{table:1}). This confirms the hypothesis that adding uncorrelated randomness on top of the topological disorder destabilizes the clean critical behavior and causes the system to flow to the generic finite-disorder fixed point of Ref.\ \cite{vojta_crewse_prb_16}. As in the case of $L_{\tau}^{\rm max}$, the values of the irrelevant exponent $\omega$ emerging from the combined fits of $m$, $\chi$, and $x_L$ do not agree well with their counterparts in the diluted square-lattice system. This can again be explained by the existence of additional irrelevant operators
\footnote{We also emphasize that extracting subleading corrections from fits is more difficult numerically than determining the leading terms, and additional uncertainties arise if more than one correction term contributes.}.

\subsection{Amplitude Mode}
\label{sec6.3}

\begin{figure}
    \centering
    \includegraphics[width=0.48\columnwidth]{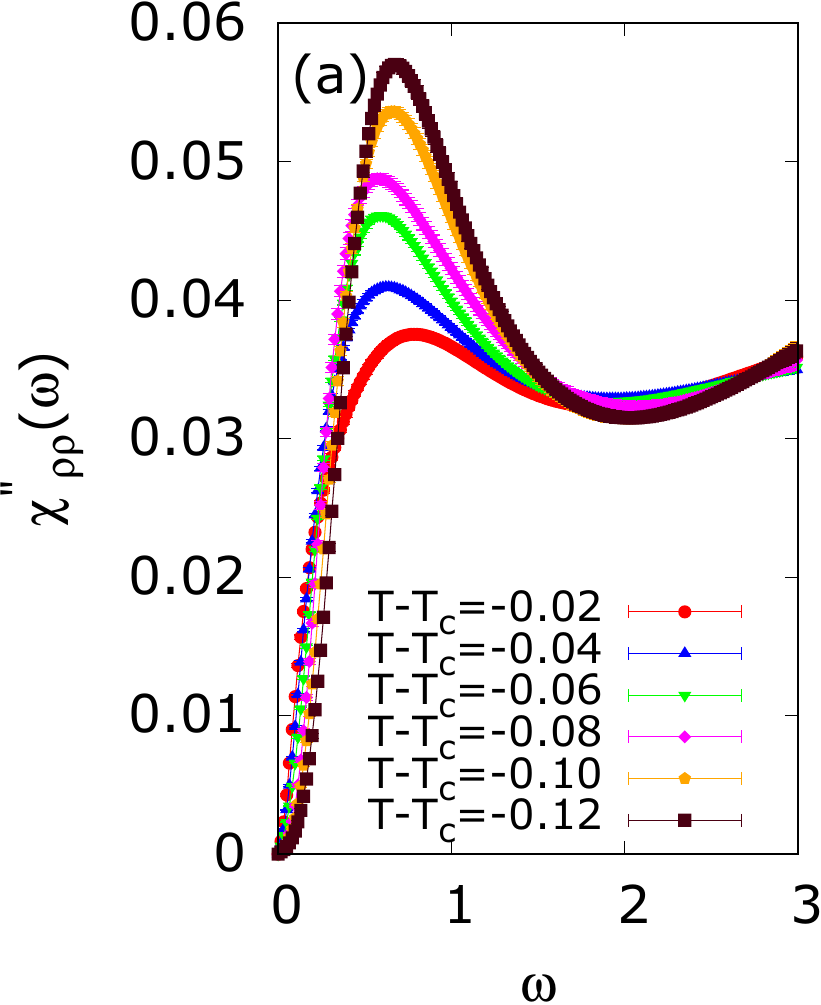}
    \includegraphics[width=0.48\columnwidth]{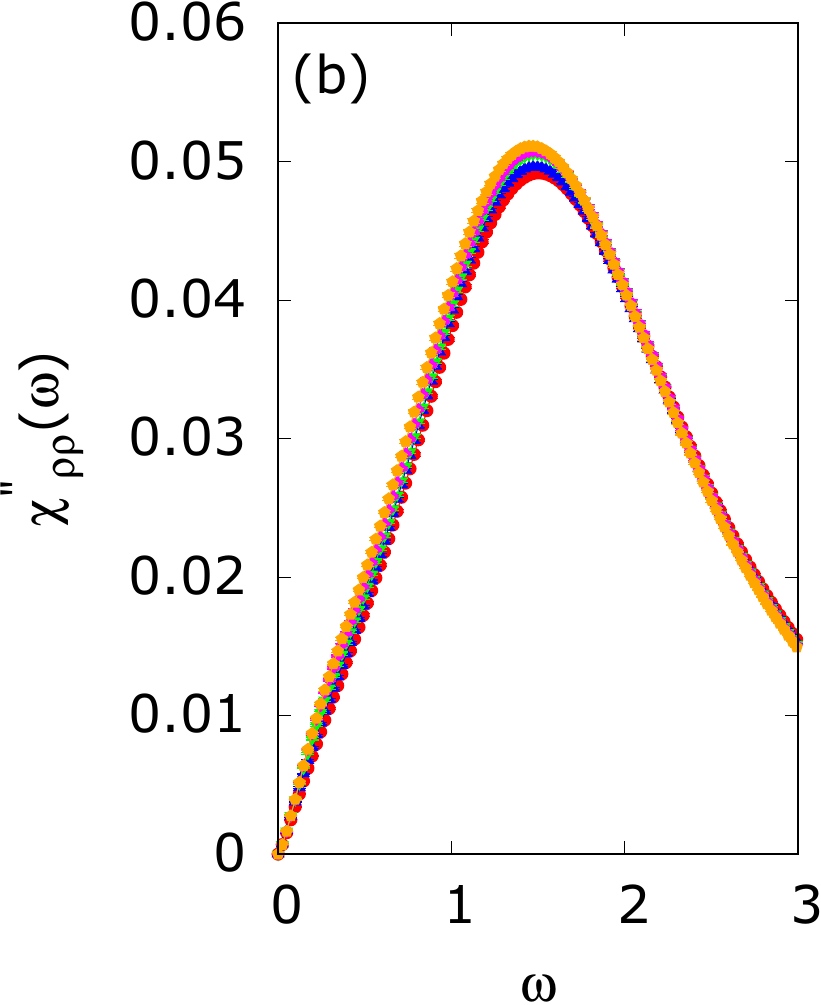}
    \caption{\justifying Spectral density $\chi^{''}_{\rho\rho}(\mathbf{q}=0,\omega)$ in the superfluid phase as a function of real frequency $\omega$ for dilutions (a) $p=0.125$ and (b) $p=0.400$. The simulations were performed at the optimal shapes (a) $L=120$ and $L_{\tau}=256$ (b) $L=88$ and $L_{\tau}=256$.}
    \label{fig:xy_generic_scaling}
\end{figure}

We now turn our attention to the behavior of the amplitude mode under the combined influence of the topological disorder of the random VD lattice and the uncorrelated disorder due to dilution. To do so, we measure the imaginary time scalar susceptibility (\ref{eqn:chirhorho}) within our Monte Carlo simulations and analytically continue it to real time and frequency by means of the maximum entropy method applied to the transformation (\ref{scaling_susceptibility}), in analogy to Sec.\ \ref{subsec:4.2}. The resulting scalar spectral functions $\chi^{''}_{\rho\rho}(\mathbf{q}=0,\omega)$ in the superfluid phase close to criticality are presented in Fig.~\ref{fig:xy_generic_scaling} for two different dilutions, $p=0.125$ and $p=0.400$.
For the weaker dilution of $p=0.125$ (Fig.~\ref{fig:xy_generic_scaling}(a)), a peak in the spectral density is still visible for the larger distances to criticality, but its magnitude is already suppressed compared to the case of pure topological disorder shown in Fig.\ \ref{fig:voronoi_peaks_without_dilution}. With decreasing distance from criticality, this peak rapidly shrinks and broadens. Moreover, the peak position does \emph{not} soften as expected from Eq.\ (\ref{eqn:scaling}) as the critical point is approached.  Thus, the scalar response of the diluted system violates naive scaling. This becomes even more obvious for the higher dilution of $p=0.400$ (Fig.~\ref{fig:xy_generic_scaling}(b)) for which the scalar spectral function does not show any low-energy peak. Instead, $\chi^{''}_{\rho\rho}(\mathbf{q}=0,\omega)$ is dominated by a broad hump centered at a microscopic energy scale. This hump is non-critical, i.e., it is completely insensitive to the distance from criticality.

\begin{figure}
    \centering
    \includegraphics[width=\columnwidth]{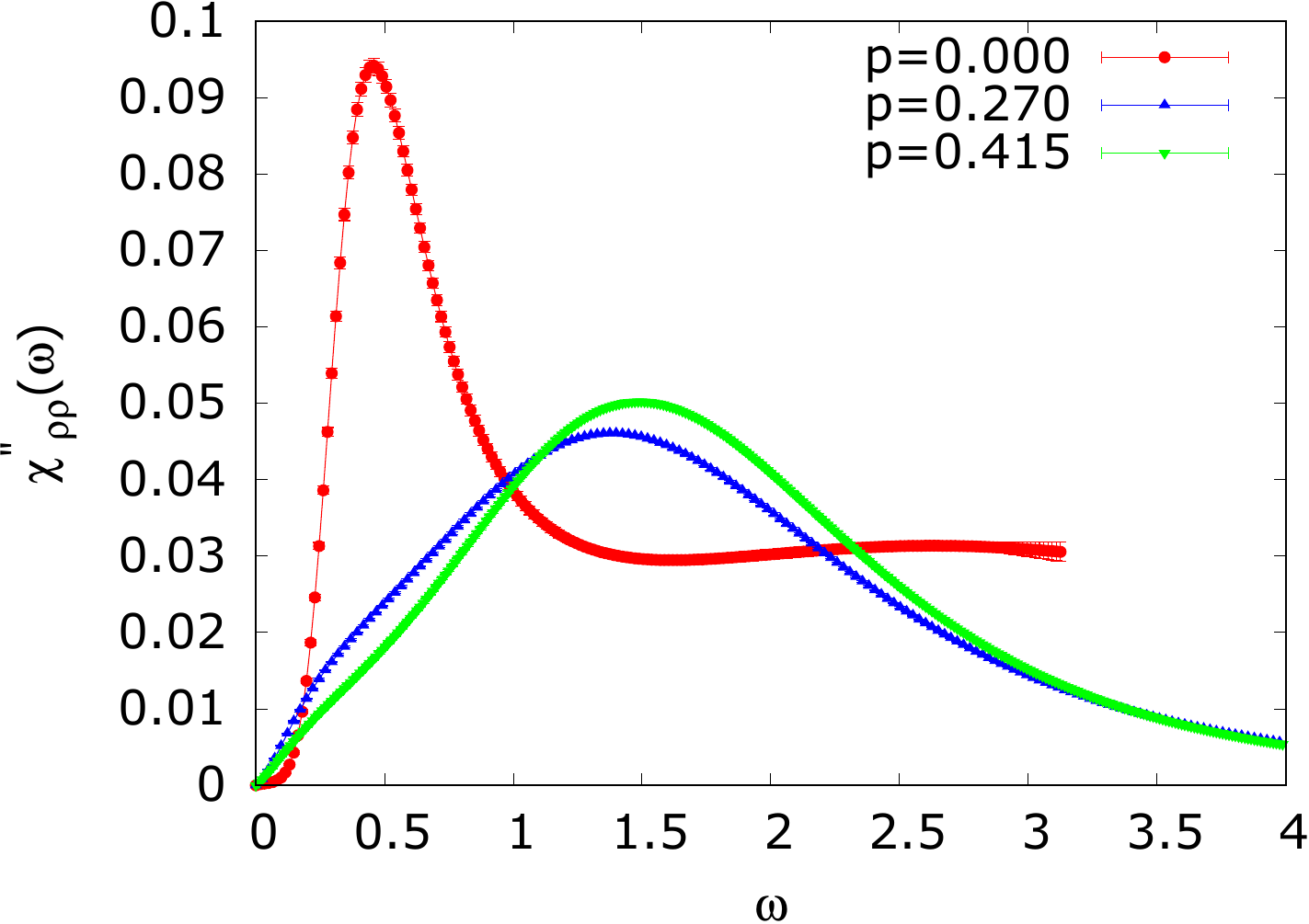}
\caption{\justifying Spectral density $\chi^{''}_{\rho\rho}(\mathbf{q}=0,\omega)$ as a function of real frequency at a fixed distance from criticality $r=|T-T_{c}|/T_{c}=-0.02$, for the dilutions $p=0$, $0.270$, and $0.415$, for $L=128$ and the optimal $L_\tau$ for each $p$.}
\label{fig:compare_dilutions}
\end{figure}

In Fig. \ref{fig:compare_dilutions}, we compare the spectral functions for several dilutions at a fixed distance from criticality, $r=-0.02$, along with the non-diluted lattice, i.e., the $p=0$ case.
For the case of the lattice with only topological disorder, we see (as in Sec.~\ref{subsec:4.2}) a sharp Higgs peak. Introducing site dilution rapidly suppresses this peak and replaces it by a broad non-critical hump.

The broadening and suppression of the Higgs peak demonstrated in Figs.~\ref{fig:xy_generic_scaling} and \ref{fig:compare_dilutions} is completely analogous to the behavior observed at the superfluid-Mott glass transition on a diluted square lattice \cite{puschmann_crewse_prl_20,crewse_vojta_prb_21}. To explain this behavior, we note that every observable close to a critical point has a singular part and a non-singular part. The singular part is governed by the long-wavelength low-energy critical fluctuations and generically follows scaling whereas the non-singular part is due to non-critical microscopic degrees of freedom. The scalar susceptibility can thus be decomposed as
\begin{equation}
    \chi_{\rho\rho}(\mathbf{q},\omega)=\chi_{\rho\rho}^{\rm reg}(\mathbf{q},\omega) + \omega^{[(d+z)\nu-2]/(\nu z)}X(\mathbf{q}r^{-\nu},\omega r^{-\nu z})~
    \label{eqn:scaling+regular}
\end{equation}
where $\chi_{\rho\rho}^{\rm reg}(\mathbf{q},\omega)$ represents the non-singular part, and the second term is the scaling form of the singular part. At the clean critical point (which also controls the transition in the system with a purely topological disorder, see Sec.\ \ref{subsec4-A}),
$z=1$ and $\nu =0.672$. Consequently, the exponent $[(d+z)\nu-2]/\nu z$ which governs the magnitude of the singular part of $\chi_{\rho\rho}$
takes a value very close to zero, meaning that this magnitude does not change as the critical point is approached. On the other hand, at the finite-disorder fixed point controlling the transitions in the diluted systems, $z=1.52$ and $\nu=1.16$. The exponent $[(d+z)\nu-2]/\nu z$ therefore takes a strongly positive value of about 1.18. This implies that the singular part is rapidly suppressed as the distance from criticality decreases, and the scalar response becomes dominated by the noncritical microscopic excitations contained in the non-singular part. For the weakest dilution we studied here, $p=0.125$, the suppression of the Higgs peak with decreasing distance from criticality can be seen directly in Fig.~\ref{fig:xy_generic_scaling}(a). For the higher dilution of $p=0.400$ shown in Fig.~\ref{fig:xy_generic_scaling}(b), the suppression is much stronger, and only the non-singular part of $\chi_{\rho\rho}$ can be seen for all temperatures.

\section{Conclusion}
\label{sec7}

In summary, we investigated the quantum critical behavior and the collective excitations near the two-dimensional superfluid-insulator quantum phase transition of interacting bosons on a random VD lattice. This was accomplished by mapping the Bose-Hubbard model onto a classical XY Hamiltonian on a (2+1)-dimensional layered VD lattice. This classical model was then simulated by means of Monte Carlo simulations to identify the universality class of the superfluid-insulator transition. We also studied the behavior of the scalar spectral function to analyze the localization properties of the amplitude mode in the presence of topological disorder. To disentangle possible mechanisms for the amplitude mode localization, the Monte Carlo simulations were complemented by an inhomogeneous mean-field theory.

Our results can be summarized as follows. In consonance with the modified Harris criterion \cite{barghathi_vojta_prl_14} (but in disagreement with the regular one \cite{Harris74}), we found that the superfluid-insulator transition in the presence of topological disorder imposed by the random VD lattice belongs to the clean 3D XY universality class. On the superfluid side of the quantum phase transition, the scalar spectral function behaves in the same way as in the clean (translationally invariant) case. Specifically, the spectral function shows a sharp peak at a Higgs energy $\omega_H$, which softens as the transition is approached and conforms to the naive scaling expectations \cite{podolsky_sachdev_prb_12,puschmann_crewse_prl_20}. This implies that the amplitude mode is not localized by the topological disorder.
As the random VD lattice is known to support Anderson localization for a system of non-interacting particles \cite{puschmann_cain_epjb_15}, our results indicate that amplitude mode localization observed in the presence of generic (uncorrelated) disorder \cite{puschmann_crewse_prl_20,crewse_vojta_prb_21}  does not stem from an Anderson-localization-type mechanism. Instead, it appears to be tied to the critical behavior of the quantum phase transition via the scale dimension of the scalar susceptibility.

This observation also agrees with our results for the Bose-Hubbard model on a random VD lattice with additional site dilution. The uncorrelated disorder introduced by the vacancies destabilizes the clean critical behavior and causes the transition to belong to the disordered universality class identified in Ref.\  \cite{vojta_crewse_prb_16}. At the same time, the sharp Higgs peak in the scalar spectral function is destroyed, and the scalar response becomes broad and non-critical.

Finally, we note that the purely topologically disordered system features remanents of a Higgs resonance even in the insulating phase close to criticality (where an order parameter can be defined on a mesoscopic length scale). This is consistent with results for the clean Bose-Hubbard model \cite{chen_liu_prl_13}.

Recently, Beattie-Hauser and Vojta \cite{BeattieHauserVojta24} studied the scalar susceptibility in a site-diluted three-dimensional classical XY model. The dilution was uncorrelated in all three directions, which means this model cannot be interpreted as arising from the quantum-to-classical mapping of a diluted quantum Hamiltonian (which would lead to disorder correlated in the imaginary time direction). Nonetheless, its scalar susceptibility can be compared with the scalar susceptibility of the mapped Hamiltonian (\ref{eqn.2}) in imaginary time (i.e., before the Wick rotation). The three-dimensional classical model with uncorrelated vacancies fulfills the conventional Harris criterion $d\nu > 2$, if just barely. Its critical behavior, therefore belongs to the clean 3D XY universality class. Beattie-Hauser and Vojta \cite{BeattieHauserVojta24} determined that the scalar susceptibility of this model in the long-range ordered phase fulfills naive scaling. This provides additional evidence that the presence or absence of unconventional behavior of the scalar susceptibility in a disordered system is controlled by its scale dimension at the transition in question.

Our results have broader implications for disordered quantum phase transitions. Based on the scaling form (\ref{eqn:scaling+regular}), one can make a general prediction about the fate of the scalar susceptibility at any quantum phase transition controlled by a finite-disorder fixed point. At such a fixed point, the correlation length exponent $\nu$ must fulfill the inequality  $d\nu > 2$ \cite{CCFS86}. Consequently, the exponent $[(d+z)\nu-2]/\nu z$ which controls the amplitude of the scalar susceptibility will be positive and larger than unity. This implies that, at any finite-disorder critical point, the singular part of scalar susceptibility is suppressed as the transition is approached, and the scalar response is dominated by non-critical microscopic fluctuations.

In many experimental systems, the (bare) disorder is expected to be only moderately strong. To interpret experimental data, it will therefore be important to understand the crossover from clean to disordered behavior as the transition is approached. This crossover is explicitly visible in Fig.~\ref{fig:xy_generic_scaling}(a), but a quantitative analysis requires the study of larger, more weakly disordered systems close to criticality. This is numerically very expensive and thus remains a task for the future.

\begin{acknowledgments}
The simulations were performed on the Pegasus and Foundry clusters at Missouri S\&T and the Aqua cluster at IIT Madras. T.V.\ acknowledges support from the National Science Foundation under Grants No.\ DMR-1828489 and OAC-1919789 as well as Grants No.\ PHY-1748958 and PHY-2309135 to the Kavli Institute for Theoretical Physics.  P.K.V.\ acknowledges support via the IIE Travel award by the Global Engagement Office, IIT Madras. R.N.\ and P.K.V.\ also acknowledge funding from the Center for Quantum Information Theory in Matter and Spacetime, IIT Madras and from the Department of Science and Technology, Govt. of India, under Grant No. DST/ICPS/QuST/Theme-3/2019/Q69, as well as support from the Mphasis F1 Foundation via the Centre for Quantum Information, Communication, and Computing (CQuICC).
\end{acknowledgments}

\section*{Appendix}
\appendix
\section{Connectivity disorder fluctuations in the random VD lattice}

The quenched disorder in the Bose-Hubbard model (\ref{eqn.1}) stems from the random connectivity of the underlying random VD lattice. At the mean-field level, the local coupling strength is proportional to the number of nearest neighbors (coordination number) that a given lattice site has. The spatial fluctuations of the coordination number of a random VD lattice were studied in detail in Ref.\ \cite{barghathi_vojta_prl_14}.
To this end, a large random VD lattice was divided into blocks of linear size $L_{b}$. (This can be done either via the real space positions of the lattice sites or via the link distance, the number of bonds separating two sites. Both partitions lead to the same results.)
A block-averaged coordination number for block no.\ $k$ can be defined by
\begin{equation}
    Q_{k}=\frac{1}{N_{b,k}}\sum_{i \in k}q_{i} ,
\end{equation}
where $q_{i}$ is the coordination number of site $i$, and $N_{b,k}$ is the number of sites in block $k$.

The spatial disorder fluctuations can be characterized by the variance of the block averaged coordination number $\sigma^2_{Q}$ and its dependence on the block size $L_b$. As the number of lattice sites in a block behaves as $L_b^d$, one might naively expect the central limit theorem result $\sigma^2_{Q} \sim L^{-d}$. However, the analysis in Ref.\ \cite{barghathi_vojta_prl_14} revealed a more rapid decrease, $\sigma^2_{Q} \sim L^{-(d+1)}$.
This rapid decay of the disorder fluctuations under coarse graining stems from a topological constraint imposed by the Euler equation of a two-dimensional graph, $N-E+F=\chi$. Here, $N$ is the number of lattice sites (vertices), $E$ is the number of edges and $F$ is the number of facets in the graph. $\chi$ is the Euler characteristic which is equal to $0$ for periodic boundary conditions (torus topology). If every facet is a triangle, as is the case for the  VD triangulation, $3F = 2E$  because each triangle has three edges, and each edge is shared by two triangles. Consequently, $E = 3N$ which implies that the total coordination does not fluctuate, and the average coordination number is exactly 6 for any
disorder realization. Thus, if there is a spatial region in the random VD lattice with above average connectivity, there must be another region with below average connectivity to compensate. The topological constraint thus introduces anti-correlations in the connectivity disorder which lead to the more rapid decay of $\sigma^2_{Q}$.

Repeating the derivation of the Harris criterion with the relation $\sigma^2_{Q} \sim L^{-(d+1)}$ rather than the central limit theorem result $\sigma^2_{Q} \sim L^{-d}$ leads to the inequality $(d+1)\nu > 2$ for the stability of a clean critical point \cite{barghathi_vojta_prl_14}.

\section{Maximum Entropy Method}

The analytic continuation of the scalar susceptibility from imaginary to real frequencies is given by Eq.\ (\ref{scaling_susceptibility}) which yields the Matsubara susceptibility $\tilde{\chi}_{\rho\rho}(i \omega_{n})$ as an integral over the real-frequency spectral function $\chi^{''}_{\rho\rho}(\omega)$. The inversion of this relation is an ill-posed problem and very sensitive to Monte Carlo noise. We can overcome this difficulty by using the maximum entropy method \cite{jarrell_guernatis_pr_96}. The idea is to use Bayesian inference to determine the most probable spectral density for a given set of Monte Carlo data for the imaginary-time susceptibility. Mathematically, the problem reduces to a minimization problem for the cost function
\begin{equation}
    Q=\Delta-\alpha S ~
\end{equation}
The first term in the cost function,
\begin{equation}
    \Delta=\left(\tilde{\chi}_{\rho\rho}-K \chi^{''}_{\rho\rho}\right)\Sigma^{-1}\left(\tilde{\chi}_{\rho\rho}-K \chi^{''}_{\rho\rho}\right)~,
\end{equation}
measures how well the Matsubara susceptibility corresponding to $\chi^{''}_{\rho\rho}(\omega)$ reproduces the input data. Here, $K$ is a discretized version of the integration kernel $2\omega/(\omega_{m}^{2}+\omega^{2})$ in Eq.\ (\ref{scaling_susceptibility}). It is given by
\begin{equation}
K(\omega,\omega_{m}) =\frac{\Delta \tau \sinh(\Delta \tau \omega)}{\cosh(\Delta \tau \omega)-\cos(\Delta \tau \omega_m)}
\end{equation}
where $\Delta \tau$ is the imaginary time step \cite{GazitPodolskyAuerbachArovas13,crewse_vojta_prb_21}.
The matrix $\Sigma_{mn}=\left<\tilde{\chi}_{\rho\rho}(i \omega_{m})\tilde{\chi}_{\rho\rho}(i \omega_{n})\right>-\left<\tilde{\chi}_{\rho\rho}(i \omega_{m})\right>\left<\tilde{\chi}_{\rho\rho}(i \omega_{n})\right>$ contains the covariance matrix elements of the scalar susceptibility Monte Carlo data.

The second term in the cost function, the entropy $S$, smoothes the resulting spectral function and prevents the overfitting of the Monte Carlo noise. It is defined as
\begin{equation}
    S=-\sum_{\omega}\chi^{''}_{\rho\rho}(\omega)\ln{\chi^{''}_{\rho\rho}(\omega)}.
\end{equation}
The free parameter $\alpha$ controls the relative weights of the two cost function terms. The optimum value of this parameter can be obtained using a particular version of the L-curve method \cite{hansen_oleary_siamjsc_93,bergeron_tremblay_pre_16}. It involves maximizing the curvature $k=d^{2}\Delta/d(\ln{\alpha})^{2}$. This maximum marks the crossover from fitting the actual information in the Monte Carlo data to fitting the noise. More details of our implementation of the maximum entropy method are described in Refs.\ \cite{puschmann_crewse_prl_20, crewse_vojta_prb_21}.

\bibliographystyle{apsrev4-2}
\bibliography{references.bib}   

\end{document}